\documentclass%[draft]
{cimento}

\def\bbox#1{\mbox{\boldmath$#1$\unboldmath}}

\title{Self--consistent Treatment of Copolymers with Arbitrary Sequences}
\author{ Yu.A.~Kuznetsov \thanks{E-mail: yuri@ucd.ie}, 
E.G.~Timoshenko\thanks{E-mail: Edward.Timoshenko@ucd.ie, Web: http://darkstar.ucd.ie}}
\instlist{
\inst{}{
Department of Chemistry, University College Dublin,
Belfield, Dublin 4, Ireland
}}
\PACSes{
\PACSit{36.20.-r}{Macromolecules and polymer molecules}
\PACSit{64.60.My}{Metastable phases}
\PACSit{64.70.Pf}{Glass transitions}
\PACSit{01.30.Cc}{Conference proceedings}
}
\begin{document}

\maketitle

\begin{abstract}

Using the Gaussian {\it Ansatz} for the monomer--monomer correlation functions
we derive a set of the self--consistent equations for determination of the
conformational state in the bead--and--spring copolymer model.
The latter is based on the Edwards type effective free energy functional
with arbitrary two--body interaction matrix.
The rate of conformational changes in
kinetics may be expressed via the instantaneous gradients of the variational
free energy functional in the space of the averaged dynamical variables.
We study the equilibrium and kinetics for some periodic and random
aperiodic amphiphilic sequences in infinitely diluted solution.
Typical equilibrium phase diagrams are elucidated and the
conformational structure of different states is discussed.
The kinetics of compaction of an amphiphilic copolymer to the globular
state proceeds through formation of locally frustrated
non--equilibrium structures.
This leads to a rather complicated multistep kinetic process.
We observe that even a small modification in the primary sequence of a
copolymer may significantly change its kinetic folding properties.

\end{abstract}

\section{Introduction}
\label{sec:intro}

Conformational transitions of heteropolymers in dilute solutions
attract special attention of theorists in recent years
\cite{BlockCopolymers}. There is a significant body of works carried out
on concentrated solutions, mixtures and blends of copolymers
based on the density variables formulation \cite{Blends}.
The infinite dilution limit, for which such approaches
cannot be used, is nevertheless very important for understanding
the fundamental interactions of macromolecules not beset with
the complications due to the aggregation phenomena.
Apart from a purely academic interest, the single chain problem
appears to be of paramount importance for building up increasingly
sophisticated models of {\it biopolymers} --- the goal that seemed
too remote only a decade ago.
The progress in the modern biotechnology is often 
impeded by the inability to solve a few fundamental theoretical problems
among which there is one grand challenge --- given the primary sequence
of macromolecule find its conformational structure at a given
equilibrium, steady, or more interestingly, nonequilibrium state.
The next part of the puzzle would be to find the relation between
the conformational structure and biofunctionality of say a protein,
but this sometimes can be explained by location of certain active sites
on the surface of a protein globule.

The computational difficulties are clearly immense for even not so
long protein chain of just about 60 amino acid residues.
Therefore at the moment one hopes to develop merely some oversimplified
models of heteropolymers abstracting from the accurate molecular level
description. Such models are obtained by coarse--graining in a fashion
normal for statistical mechanics by integrating out many of the 
local degrees of freedom. We hope to be able to deduce the mesostructure
of a macromolecule and to obtain many of the important observables.
The approach that we employ here is in its essence a nonequilibrium
extension of the Gibbs--Bogoliubov variational principle.
The main strength of the Gaussian self--consistent (GSC) method,
which we \cite{GscKinet,GscHomKin,GscBlock} and others \cite{Allegra}
were developing first for the simple homopolymer and in Ref.~\cite{GscArbCop}
we have brought to the most general form capable of describing copolymers,
is in that it produces the complete set of mean squared distances
between pairs of monomers and thus the conformational structure.

In this work we do not concentrate on the formal derivation of
the GSC equations. This is presented in some detail in Ref. \cite{GscArbCop}.
First, we shall introduce a crucially important modification
to the model itself by adding a new so--called self--interaction term.
We prove that without this regularising term the theory is plagued
with divergences if at least one second virial coefficient is negative.
We show that although the new term is seemingly negligible for long
chain lengths, and indeed is not required for the coil,
it does in fact the trick of correcting the structure
of the dense globular state. 
In the rest of the work 
we carry out extensive study of the equilibrium properties
and the folding kinetics for a number of examples of heteropolymer sequences.
We also discuss how the spin glass behaviour arises for sufficiently
random copolymer sequences.

\section{The Model and the GSC Equations}
\label{sec:model}

The main variables in the coarse--grained description of the polymer
chain \cite{Edwards,deGennes,Cloizeaux} are the spatial monomer
coordinates ${\bf X}_n$, where $n$ is the monomer number.
The solvent molecules are excluded from the consideration by integrating
out their degrees of freedom from the path integral representation for
the partition function.
The resulting monomer interactions are represented by the effective free
energy functional (EFEF),
\begin{equation} \label{eq:H}
H= \frac{k_B T}{2 l^2}\sum_n ({\bf X}_n-{\bf X}_{n-1})^2+
 \sum_{J=2}^{\infty}\sum_{\{n\}}u^{(J)}_{\{n\}} \prod_{i=1}^{J-1}
 \delta({\bf X}_{n_{i+1}}-{\bf X}_{n_1}),
\end{equation}
where for heteropolymers $u^{(J)}_{\{n\}}$ are in principle allowed to have
any dependence on the site indices $\{n\}\equiv \{n_1, \ldots, n_J\}$.
The first term in Eq. (\ref{eq:H}) describes the connectivity of the
chain with $l$ called the statistical segment length.
There are also volume interactions represented by the
virial--type expansion \cite{deGennes,Cloizeaux} in Eq. (\ref{eq:H}).
They reflect the hard--core repulsion, the weak attraction between monomers
and the effective interaction mediated by the solvent--monomer couplings.

Here we consider the following choice of site--dependent second
virial coefficients in Eq. (\ref{eq:H}),
\begin{equation} \label{eq:u2}
u^{(2)}_{nn'}=\bar{u}^{(2)}+\Delta\frac{\sigma_n+\sigma_{n'}}{2},
\end{equation} 
and $u^{(J)}_{\{ n \} } = u^{(J)}$ for $J > 2$.
This corresponds to the case of amphiphilic heteropolymers, for which
monomers differ only in the monomer--solvent coupling constants.
The {\it mean second virial coefficient}, $\bar{u}^{(2)}$, is associated
with the quality of the solvent and the parameter $\Delta$ is called the
{\it degree of amphiphilicity} of the chain.   
The set $\{\sigma_n\}$ expresses the chemical composition, or the
{\it primary sequence} of a heteropolymer. Here we consider the case when
variables $\sigma_n$ can take only three values:
$-1,1$ and $0$ corresponding to hydrophobic '$a$', hydrophilic '$b$' and 
``neutral'' '$c$' monomers respectively.

It is assumed that the long timescale evolution of the conformational state
is well represented by the Langevin equation, which upon neglecting the
hydrodynamics may be written as,
\begin{equation} \label{eq:Lang}
\zeta_b \frac{d}{dt} {\bf X}_n = - \frac{\partial H}{\partial {\bf X}_n}
+\bbox{\eta}_n(t),
\end{equation}
where $\zeta_b$ is the ``bare'' friction constant per monomer and
the Gaussian noise, $\bbox{\eta}_n$, is characterised by the second momentum,
\begin{equation} \label{eq:noise}
\langle \eta^{\alpha}_n(t)\,\eta^{\alpha'}_{n'}(t') \rangle
=2k_B T \zeta_b\, \delta^{\alpha,\alpha'}\delta_{n,n'}\delta(t-t'),
\end{equation}
where the Greek indices denote the spatial components of 3-d vectors. 

The main idea of the Gaussian self--consistent method is to choose the
trial Hamiltonian, $H_0$, as a most generic quadratic form, with matrix
coefficients depending on time,
\begin{equation} \label{eq:trial}
H_0 (t) = \frac{1}{2}\sum_{nn'} V_{nn'}(t)\,{\bf X}_n(t)\,{\bf X}_{n'} (t). 
\end{equation}
This corresponds to the Gaussian distribution of the inter--monomer
distances, $({\bf X}_m - {\bf X}_{m'})^2$.
Thus, the two--body monomer--monomer correlation function,
that is the probability density to find the monomer $m'$ at
the distance $r$ from the monomer $m$, will be given by,
\begin{equation} \label{eq:corr2}
h^{(2)}_{mm'} ({\bf r};\,t) \equiv
\left\langle \delta({\bf r} - {\bf X}_m + {\bf X}_{m'})\right\rangle =
\frac{1}{(2\pi D_{mm'}(t) )^{3/2}}\exp\left(-\frac{r^2}{2 D_{mm'}(t)}\right).
\end{equation}
Here $D_{mm'}$ is the matrix of the mean squared distances between monomers,
\begin{equation}\label{eq:Dmn}
D_{m\,m'}(t) \equiv \frac{1}{3}\left\langle ({\bf X}_m(t) 
- {\bf X}_{m'}(t))^2 \right\rangle.
\end{equation}
Obviously, choosing Eq. (\ref{eq:trial}) as the trial Hamiltonian is
equivalent to replacing the nonlinear Langevin equation (\ref{eq:Lang})
by a linear stochastic ensemble,
\begin{equation} \label{eq:LL}
\zeta_b \frac{d}{dt}{\bf X}_n =- \sum_{n'} V_{nn'}(t)\,{\bf X}_{n'} +
\bbox{\eta}_n(t).
\end{equation}
The time--dependent coefficients are chosen at each moment in time
according to the criterion,
\begin{equation} \label{eq:crit}
\left\langle {\bf X}_n\, \frac{\partial H}{\partial {\bf X}_{n'}} 
\right\rangle_0 = \left\langle {\bf X}_n\, 
\frac{\partial H_0}{\partial {\bf X}_{n'}} \right\rangle_0, \qquad
\end{equation}
where $\langle\ldots\rangle_0$ denotes the averaging over the trial ensemble.
At equilibrium these equations become exactly the extrema conditions 
for the trial free energy in the Gibbs--Bogoliubov variational principle
based on minimising the variational free energy, ${\cal A}
=-k_B T\log\mbox{Tr} \exp(-H_0/k_B T)
+\langle H-H_0\rangle_0$, with respect to $V_{nn'}$.

For details of calculations we refer the reader to Refs.
\cite{GscBlock,GscArbCop}. Here we present the final form of the kinetic
GSC equations which describe the time evolution of the mean
squared distances between monomers (\ref{eq:Dmn}).
It turns out that the equations can be written in terms of instantaneous
gradients of the variational free energy, ${\cal A} = {\cal E}-T{\cal S}$,
\begin{equation} \label{eq:kinmain}
\frac{\zeta_b}{2}\frac{d}{dt}D_{mm'}(t) = -\frac{2}{3}
\sum_{m''}(D_{mm''}(t)-D_{m'm''}(t))\left(
\frac{\partial {\cal A}}{\partial D_{mm''}(t)}-
\frac{\partial {\cal A}}{\partial D_{m'm''}(t)}
\right).
\end{equation}
The energetic and the entropic contributions in the free energy can be
completely expressed in terms of the mean squared distances $D_{mm'} (t)$,
\begin{eqnarray}
{\cal E} &=& \frac{3k_B T}{2l^2}\sum_n D_{n\,n-1,n\,n-1}
    +\sum_{J=2}^{\infty}\sum_{\{n\}'}\frac{u^{(J)}_{\{n\}}}{(2\pi)^{3(J-1)/2}}
    ({\rm det} \Delta^{(J-1)})^{-3/2} + {\cal E}_{si}, \label{eq:E}\\
{\cal S} &=& \frac{3}{2}k_B \log {\rm det}\, R^{(N-1)},\qquad
    R_{nn'}=\frac{1}{N^2}\sum_{mm'}D_{nm,n'm'}, \label{eq:S}
\end{eqnarray}
where we have introduced the four--point correlation function and the matrix
$\Delta^{(J-1)}$,
\begin{eqnarray}
D_{mm',nn'} &\equiv& \frac{1}{2}(D_{m'n}+D_{mn'}
-D_{mn}-D_{m'n'}), \label{D4} \\
\Delta^{(J-1)}_{ij} &\equiv& D_{n_1 n_{i+1},n_1 n_{j+1}}.
\end{eqnarray}
In Eq. (\ref{eq:S}) we have the determinant of the truncated matrix
$R^{(N-1)}$ to exclude the zero eigenvalue related to
the translational invariance for the centre--of--mass of the system.
In the second term in Eq. (\ref{eq:E}), which is responsible for the volume
interactions, the summation is taken over not coinciding indices,
$n_1 \not= n_2 \not= \ldots \not= n_J$.

Before proceeding with further discussions of the GSC equations
let us introduce some observables. These include
the mean squared radius of gyration,
\begin{equation} \label{eq:rg2}
R_g^2 = \frac{1}{2 N^2}\sum_{mm'} D_{mm'},
\end{equation}
and the micro--phase separation (MPS) order parameter,
\begin{equation} \label{eq:mps}
\Psi  = \frac{1}{N^2\,R_g^2}
\sum_{mm'}\frac{\sigma_m + \sigma_{m'}}{2 \Delta_\sigma} D_{mm'},\qquad
(\Delta_\sigma)^2 = \frac{1}{N}\sum_m \sigma_m^2.
\end{equation}
The MPS parameter describes the degree of correlation between
matrices of the relative two--body interaction,
$(\sigma_m + \sigma_{m'})/2$, and the mean squared distances, $D_{mm'}$.

\section{The Self--interaction Energy Term}
\label{sec:si}

The appearance of the last term in Eq. (\ref{eq:E}), ${\cal E}_{si}$,
is somewhat more nontrivial.
In fact, in the EFEF of the model (\ref{eq:H}) we have discarded
terms with two or more coinciding indices in the three-- and higher body
contributions. These terms come formally from the virial--type expansion,
but each of them gives a singular contribution to the mean energy (\ref{eq:E}).
It turns out, however, that upon suppressing these terms
there appear additional pathological solutions of the GSC equations
with singular free energy if at least one element of the
two--body interaction matrix, $u^{(2)}_{mm'}$, becomes negative.
This is easy to see. Indeed,
consider volume interactions of just three monomers under condition
that  the mean squared distances from monomers '0' and '1' to '2' are
equal to each other, $D_{0,2} = D_{1,2} = D$. These interactions
produce the mean energy contribution, 
\begin{equation} \label{eq:nons}
{\cal E}_3 = \frac{u^{(2)}_{0,2} + u^{(2)}_{1,2}}{(2\pi D)^{3/2}} +
             \frac{1}{(2\pi D_{0,1})^{3/2}} \left( u^{(2)}_{0,1} +
             \frac{6 u^{(3)}(2\pi)^{-3/2}}{(D - D_{0,1}/4)^{3/2}} \right)
\end{equation}
In the case when $u^{(2)}_{0,1} < 0$ and monomer '2'
is placed away from monomers '0' and '1',
$D > D_{0,1}/4 + (6 u^{(3)}(2\pi)^{-3/2}/|u^{(2)}_{0,1}|)^{2/3}$,
obviously in the limit $D_{0,1} \rightarrow 0$ the
energy possesses a singular minimum, ${\cal E}_3 \rightarrow -\infty$.
As for the free energy, the logarithmic divergence of the entropy could
not change the situation, thus ${\cal A} \rightarrow -\infty$ as well.
One can show that the inclusion of more monomers in the chain or of higher
than three--body interactions does not improve the situation, but produces
more and more of such pathological solutions.
The reason we have not discussed this problem in our previous considerations
is that we have accounted for the additional symmetry properties of
monomer--monomer distances, which come from the symmetry of the EFEF
(\ref{eq:H}). For example, in the case of the ring homopolymer, due to the
inverse symmetry
\cite{GscBlock}, we assumed that for any indices $m$, $m'$ the following
mean squared distances are equal, $D_{m,m'} = D_{m,2m-m'} = D_{2m'-m,m'}$.
This provides sufficient repulsion coming from three--body term to preclude
pathological solutions.

Thus, in a more general case, where no symmetry properties could be assumed
for an arbitrary sequence, the standard procedure of suppressing terms
with coinciding indices is not satisfactory. Fortunately, it could
be remedied by using another prescription --- replacing the terms with
coinciding indices by the so--called self--interaction terms.
Here we propose the prescription for three--body interaction which is
sufficient for our current purposes,
\begin{equation} \label{eq:Esi}
{\cal E}^{(3)}_{si} = c_3 u^{(3)}\sum_{m\not=m'}\biggl\langle
 \delta({\bf X}_{m}-{\bf X}_{m'}) \biggr\rangle^2 =
 c_3 \hat{u}^{(3)}\sum_{m\not=m'} D_{mm'}^{-3},
\end{equation}
where $c_3 = 3$ is a combinatorial factor related to the three possible
ways of having coinciding pairs of indices in a triple summation.
Obviously, the higher negative power of $D_{mm'}$ in (\ref{eq:Esi}) compared
to the two--body term in (\ref{eq:E}) prevents
one monomer from falling on another.

It would be interesting to consider the coefficient $c_3$ in
Eq.~(\ref{eq:Esi}) as an independent parameter and to discuss how
the inclusion of this term would change the equilibrium and kinetics
for the ring homopolymer \cite{Footnote0}.
Note that for the ring homopolymer we can reduce the number of independent
elements in the matrix of mean squared distances, $D_{mm'}$, by the factor
of $N$, since due to the translational symmetry,
$D_{m,\,m'} = D_{m+k,\,m'+k}$ for any $k$ and $m$, $m'$.
This symmetry allows us to reduce significantly the computational requirements.

In the good solvent regime, $u^{(2)} > u^{(3)} > 0$, we find no substantial
change caused by the self--interaction term (\ref{eq:Esi}). The deviations
in the mean squared distances and in the squared radius of gyration
typically are less than $1\%$ even for short chains.
This is only natural.
In this regime this term is subdominant for long chains and
can be neglected at all in the thermodynamic limit.

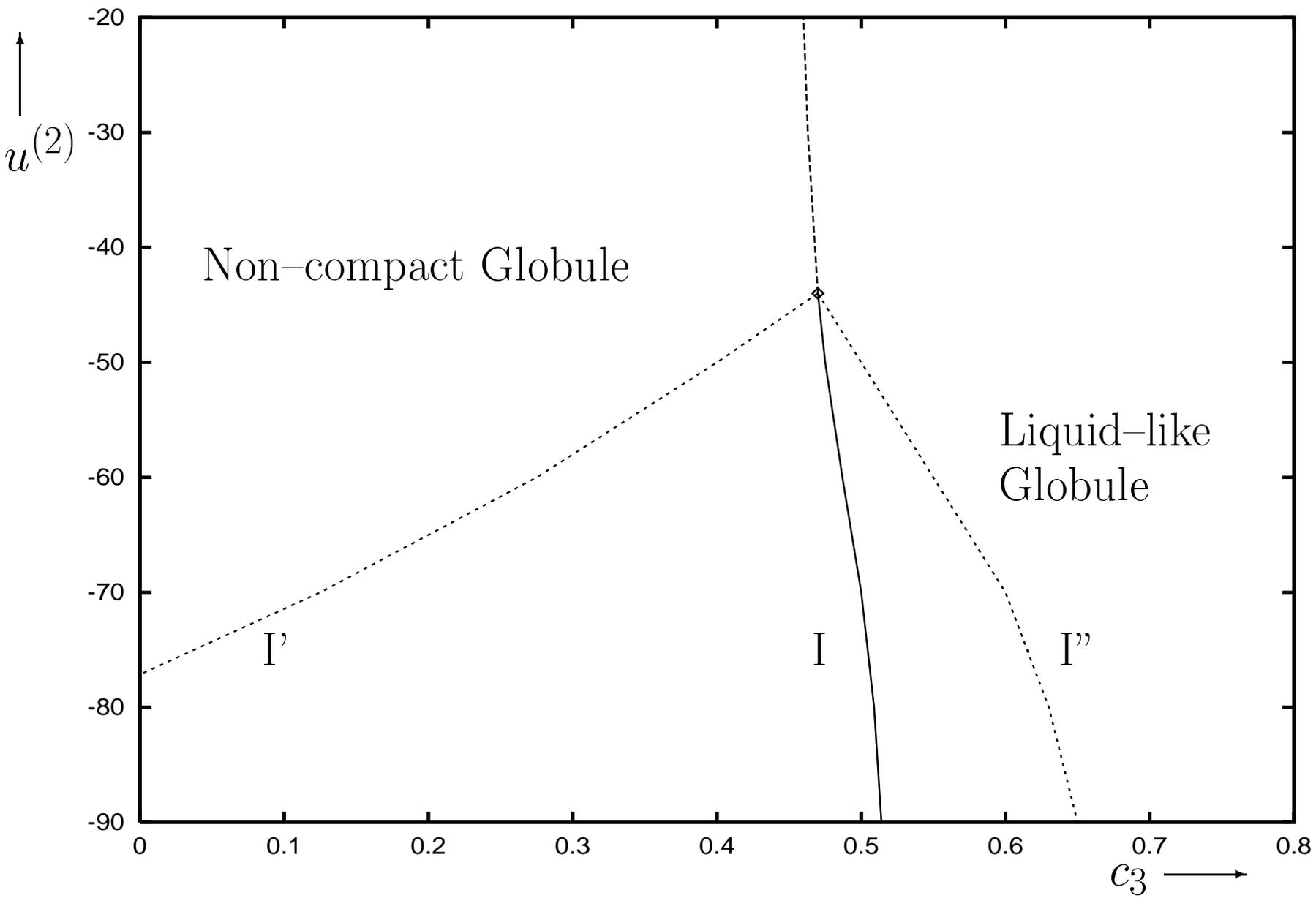
\begin{figure}
\caption{
Diagram of stability of the liquid--like versus the non--compact globules
in term of the second virial coefficient, $u^{(2)}$, and the self--interaction
parameter, $c_3$. Curve I corresponds to discontinuous transition,
curves I' and I'' are spinodals.
}
\label{fig:si}
\end{figure}

However, the size and the structure of the homopolymer globule, which
exists in the region $u^{(2)} < 0$, can change significantly.
In Fig. \ref{fig:si} we exhibit the diagram of globular states.
One can see that actually two different globular states are possible,
which we call the non--compact globule (as one can see
from Table \ref{tab:si} this globular state possesses a higher value of
$R_{g}^2$) and the liquid--like globule.
The former is the thermodynamically stable state at comparatively
small $c_3$, whilst the latter exists at higher values of $c_3$ and in the
region of large $u^{(2)}$. For large negative $u^{(2)}$ the transition
between two globular states becomes discontinuous and the transition
line goes nearly vertically at approximately $c_3 \approx 1/2$.
However, the collapse transition from the extended polymer coil to the
liquid--like globule remains second order. We found that this
phase diagram is quite independent of the degree of polymerization, at
least in the range we could study numerically, $30 \leq N \leq 200$.
Thus, introducing of the term (\ref{eq:Esi}), which is subdominant for
large values of $N$ compared to other terms in (\ref{eq:E}), nevertheless
dramatically changes the properties of the globular state even for very
long chains.

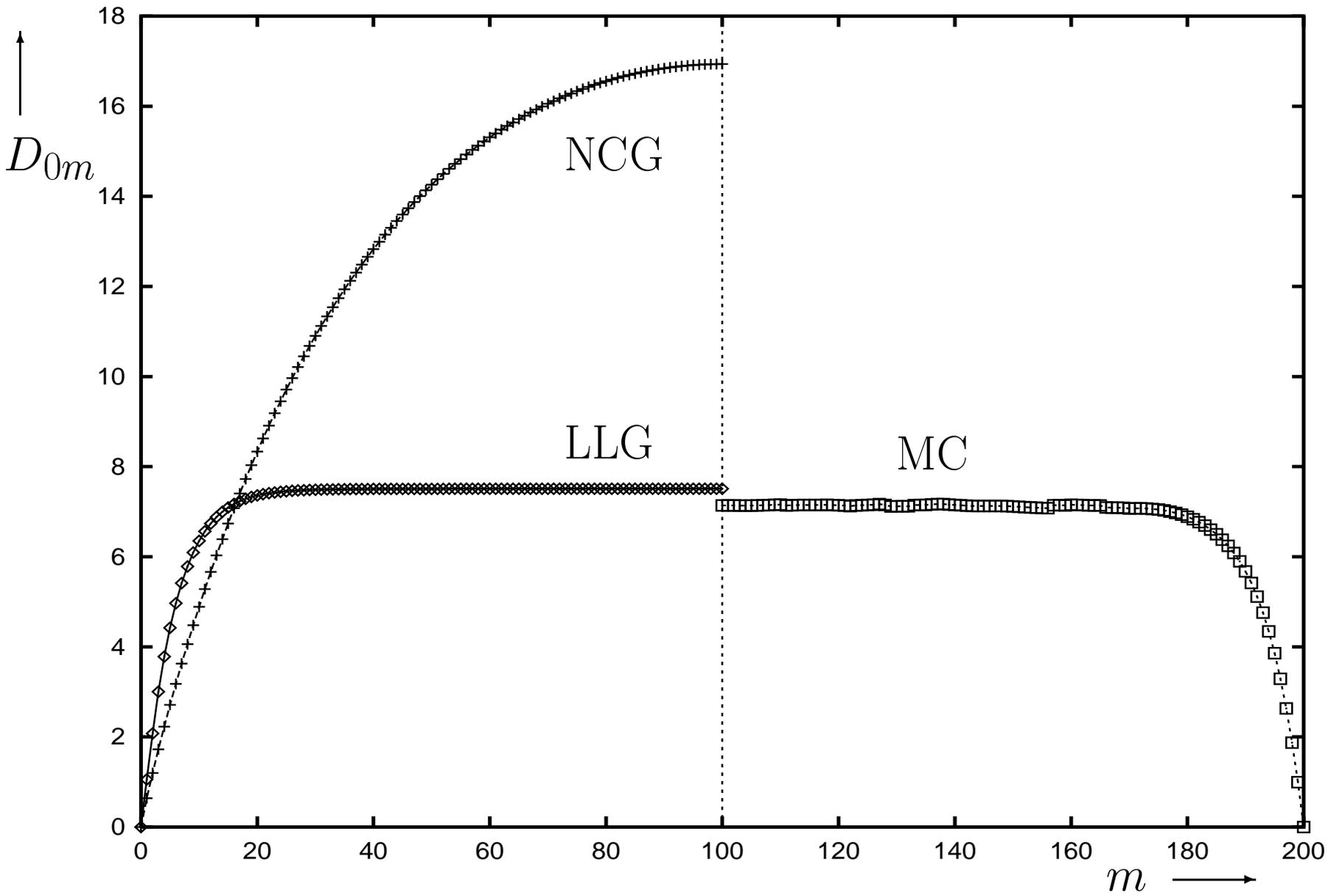
\begin{figure}
\caption{
Plot of the mean squared distances between monomers, $D_{0m}$,
versus the chain index, $m$, for polymer with $N = 200$ for the
non--compact (NCG), $c_3 = 0$, and the liquid--like (LLG) globules, $c_3 = 3$.
Here $u^{(2)} = -25$ and $u^{(3)} = 10$.
In the right--hand side we present the data from Monte Carlo simulations
\cite{CMCUnp} for the same polymer size.
Since the function $D_{0m}$ is symmetric with respect to the value
$N/2$ we present these dependencies only on half--interval.
}
\label{fig:dmh}
\end{figure}

Now let us compare the structure of the non--compact
and the liquid--like globules.
In Fig. \ref{fig:dmh} in the left--hand side we present the mean squared
distances, which are symmetrical with respect to the line $m=N/2$, i.e.
$D_{0,|m-m'|} = D_{0,N-|m-m'|}$, for values of $c_3 = 0,\ 3$.
In the right--hand side of the figure we draw the same quantity obtained
from Monte Carlo simulations for a ring chain of hard spheres with
the Lennard--Jones attraction \cite{TorNew}.
Since the parameters of excluded volume interactions in the model used
for Monte Carlo simulations are expressed in different terms,
only shape and scaling behaviour, but not the absolute values of each
$D_{mm'}$, should be compared.
We see a quite remarkable agreement here.
This is a strong argument in favour of the liquid--like globule.
Its mean squared distances have a typical saturation regime after
comparatively small $m$. This is known to be true of the globule from
Monte Carlo simulations and previous results from other methods
\cite{GrosbKhokh}.
The function, $D_{0m}$ for the non--compact globule has much more convex
shape which, in fact, reflects the effective monomer repulsion on
large distances (see Eq. (\ref{eq:nons}).
Also, from Table \ref{tab:si} we can see that the asymptotic scaling
in the degree of polymerisation for the size of the liquid--like globule,
$R_g^2 \sim N^{2/3}$, is reached starting from sufficiently
small $N$ in agreement with lattice Monte Carlo simulations
\cite{CoplmMonte}, whilst from much larger $N$ for the non--compact
globule \cite{GscHomKin}.

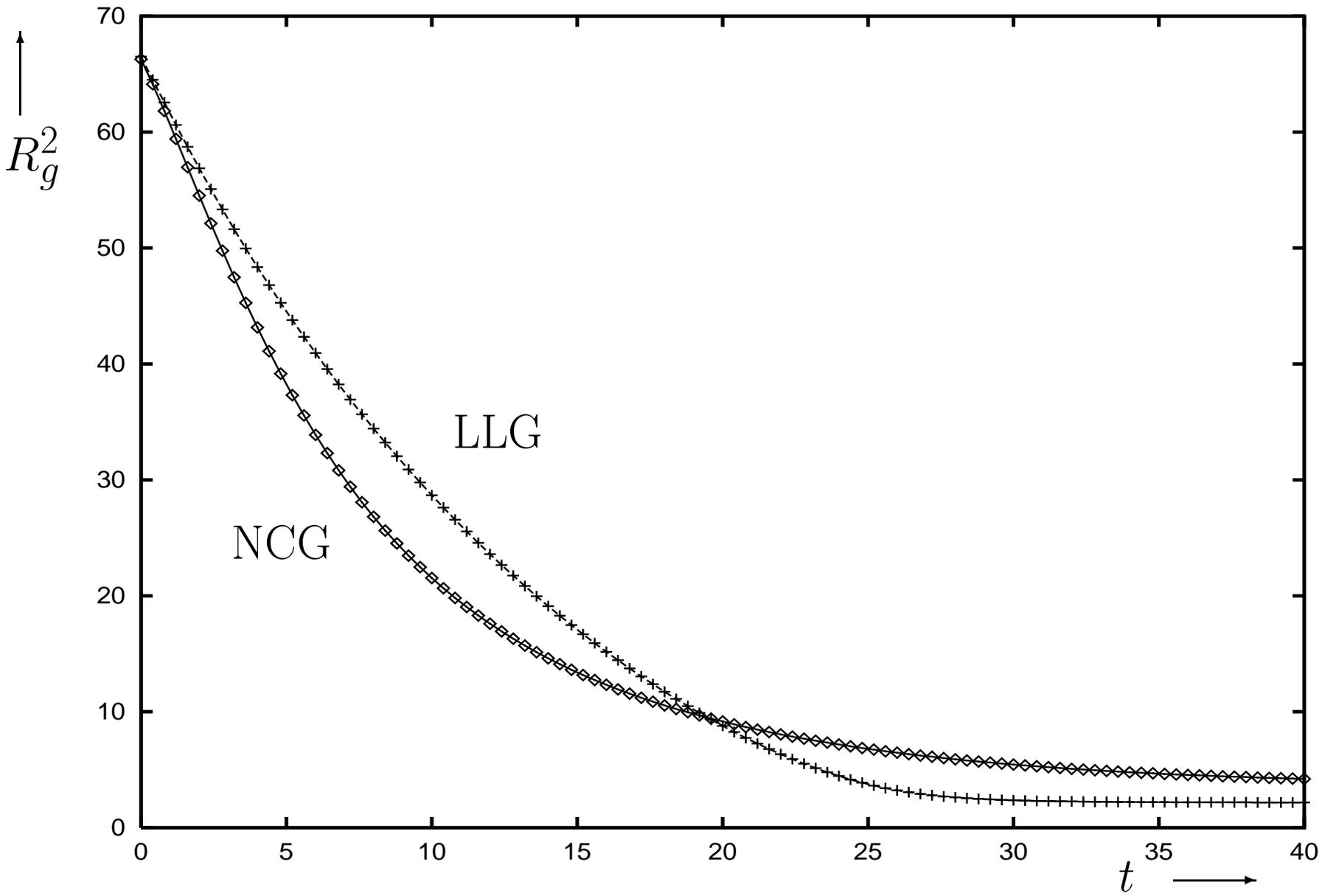
\begin{figure}
\caption{
Plot of the mean squared radius of gyration, $R_g^2$ versus time, $t$,
at the kinetics of the collapse transition, $u^{(2)}=15 \rightarrow -25$
to the non--compact (NCG), $c_3=0$, and the liquid--like (LLG), $c_3=3$,
globules.
}
\label{fig:kinsi}
\end{figure}

Now, let us turn our attention to the kinetics at the collapse
transition, after an instantaneous change of the value of the
second virial coefficient from a positive to a negative value, the latter
corresponding to the globular equilibrium state.
In Fig. \ref{fig:kinsi} we exhibit the evolution of the mean squared radii
of gyration after quenches to the liquid--like and the non--compact globule.
We found that the early stage proceeds in a similar manner, while
the middle or ``coarsening'' kinetic stage is somewhat slower for
quenches to the liquid--like globule, but the final stages here
are much faster.
This can be seen from Tab. \ref{tab:si}, where we present the
values of the total collapse time \cite{Footnote1} and the final
relaxation time, $\tau_f$ for different sizes, $N$.
The most striking thing here is that the effective exponent of the final
relaxation time is much smaller than what we have earlier expected,
$\gamma_f = 5/3$ (see Ref. \cite{GscKinet,GscHomKin}).
This also shifts down the effective exponent of the ``total''
collapse time, since the latter is a cross--over between
the exponent of the coarsening stage, $\gamma_m = 2$,
which is unaffected by the self--interaction term,
and the exponent of the final relaxation, $\gamma_f$.
In principle, the final relaxation time and its exponent in the degree
of polymerization can be determined without appealing to numerical
solution of the kinetic GSC equations. From Ref. \cite{GscKinet} we have,
$\tau_f \sim N {\cal F}_1$, where ${\cal F}_1$ is the first normal mode
for the final equilibrium globular state,
\begin{equation} \label{eq:Fq}
{\cal F}_q = -\frac{1}{2N}\sum_{m} \cos\frac{2\pi qm}{N} D_{0m}.
\end{equation}
A good approximation for the function $D_{0m}$ in the liquid--like globule
would be the following $D_{0m}$ is linearly increasing function of $m$
until some value $D_{0m_1} = D$, where the $N$-dependence is,
$D\sim m_1 \sim N^{2/3}$ and then it remains constant. This approximation
can also be obtained from the Lifshitz theory \cite{GrosbKhokh}.
Now one can see that ${\cal F}_1 \sim N^{1/3}$
and, correspondingly, $\gamma_f = 4/3$, which is close to
the value of the effective exponent in Tab. \ref{tab:si}.
Thus, on the contrary to the coil state, for the liquid--like globule
the first normal mode, ${\cal F}_1$, neither gives the main
contribution to the mean squared radius of gyration, nor even has
the same scaling law in the degree of polymerization, $N$.

Inclusion of the self--interaction term (\ref{eq:Esi}) also strongly
affects the phase diagram and kinetic properties of rigid chains.
We shall consider these questions in a separate work \cite{TorNew}.
Such modifications are rather welcome and allows one to make the GSC
method more accurate for the dense globular states,
where its validity has been less established.

\section{Equilibrium Properties of Copolymer Sequences}
\label{sec:eq}

The GSC equations (\ref{eq:kinmain}) have been studied numerically
using the fifth order Runge--Kutta algorithm with adaptive
time step \cite{NumerRecip}. In fact, for copolymer sequences, considered
in the current work, the time step during numerical integrations of
Eqs.~(\ref{eq:kinmain}) varies approximately 100 times. Thus, any
integration scheme with fixed time step is rather unreliable here.

In studying the equilibrium we consider only stationary points of
Eqs. (\ref{eq:kinmain}), i.e. the limit $t\rightarrow\infty$.
If for some set of interaction parameters, $\bar{u}^{(2)}$
and $\Delta$, one obtains several stationary states, one should compare the
values of the variational free energy, ${\cal A}$. The deepest minimum of
the free energy corresponds to the thermodynamically stable state, the rest
of the solutions to metastable states.
Here we consider copolymers with the ring topology, though the current
treatment may be easily extended for study of copolymers with any other
topology just by changing the spring term in Eqs. (\ref{eq:H}, \ref{eq:E}).

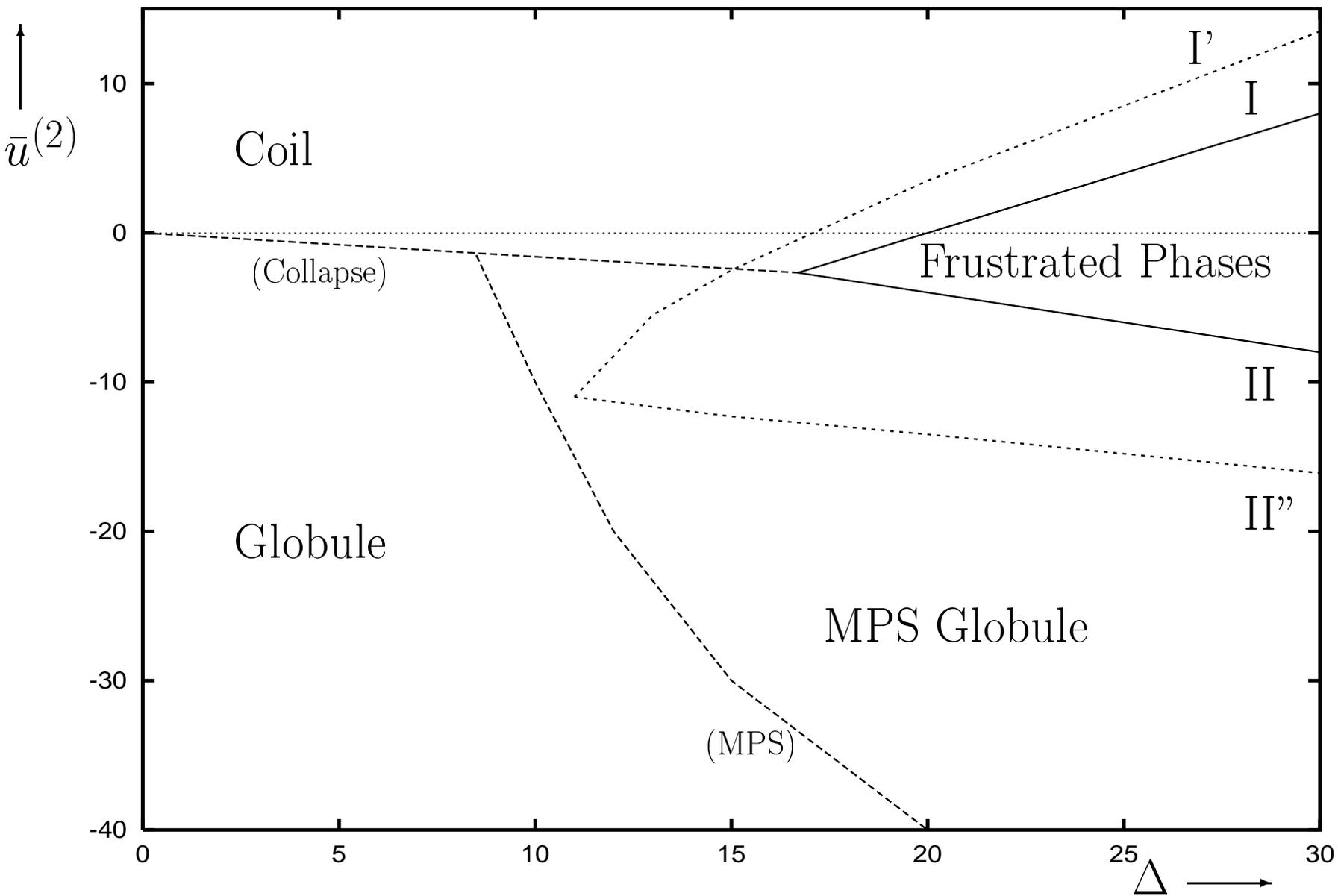
\begin{figure}
\caption{
The phase diagram of ``random'' sequence
'$babca_2cbac_2acb_3cac_3a_2b_2cac_2b$'
in terms of the mean second virial coefficient, $\bar{u}^{(2)}$, and the
amphiphilicity, $\Delta$. Curves (Collapse) and (MPS) correspond respectively
to the collapse and the micro--phase separation continuous transitions.
Curves (I) and (II) correspond to discontinuous transitions to the
frustrated phases. ``Spinodal'' curves (I') and (II'') bound the regions
of metastability of the frustrated states. Transition curves and boundaries 
distinguishing different frustrated states are not depicted.
}
\label{fig:ran30}
\end{figure}

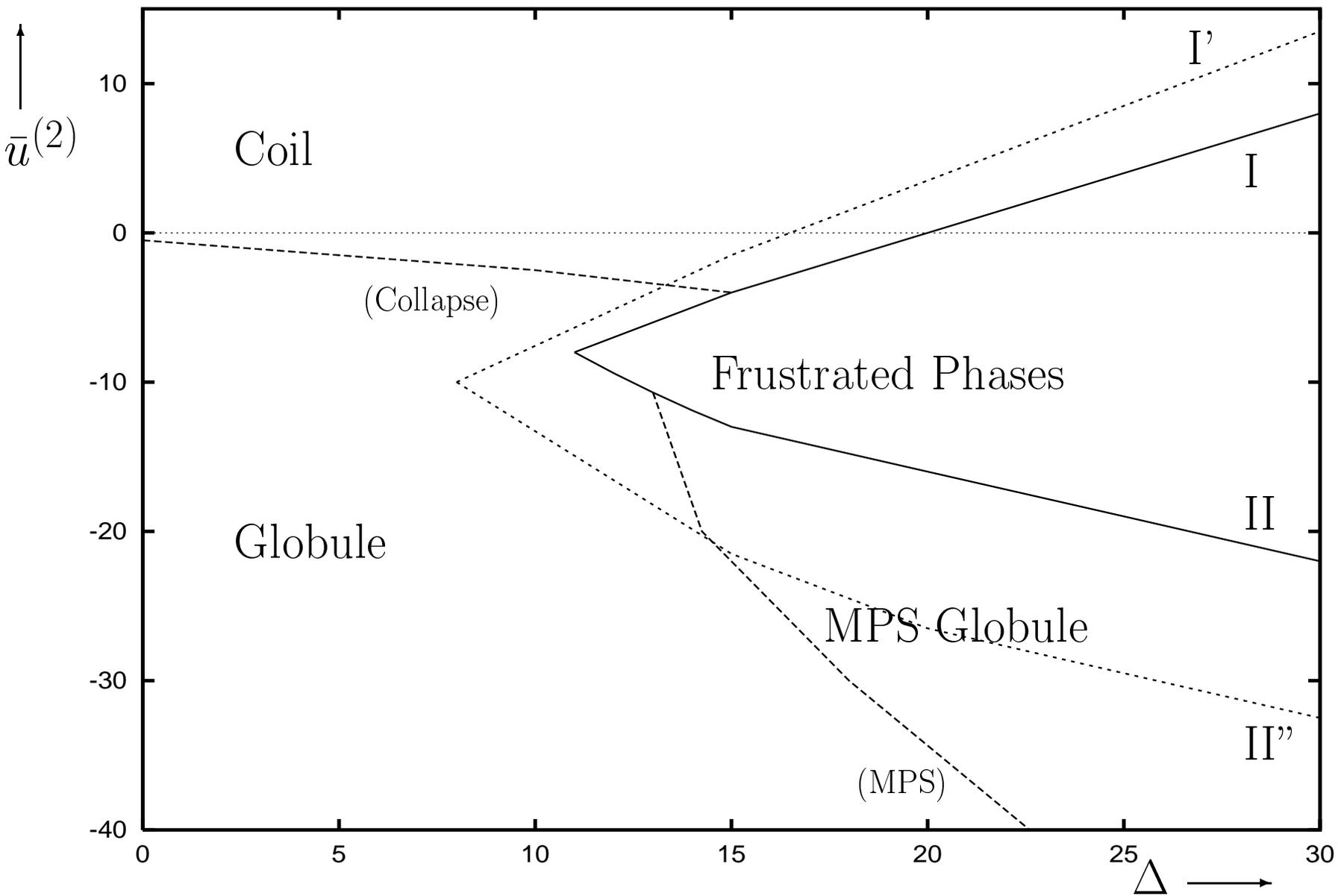
\begin{figure}
\caption{
The phase diagram of ``random'' sequence
'$(babca_2cbac_2acb_3cac_3a_2b_2cac_2b)_2$', i.e. twice as in
Fig. \ref{fig:ran30} in terms of the mean second virial coefficient,
$\bar{u}^{(2)}$, and the amphiphilicity, $\Delta$.
See caption to Fig. \ref{fig:ran30} for more details.
}
\label{fig:ran60}
\end{figure}

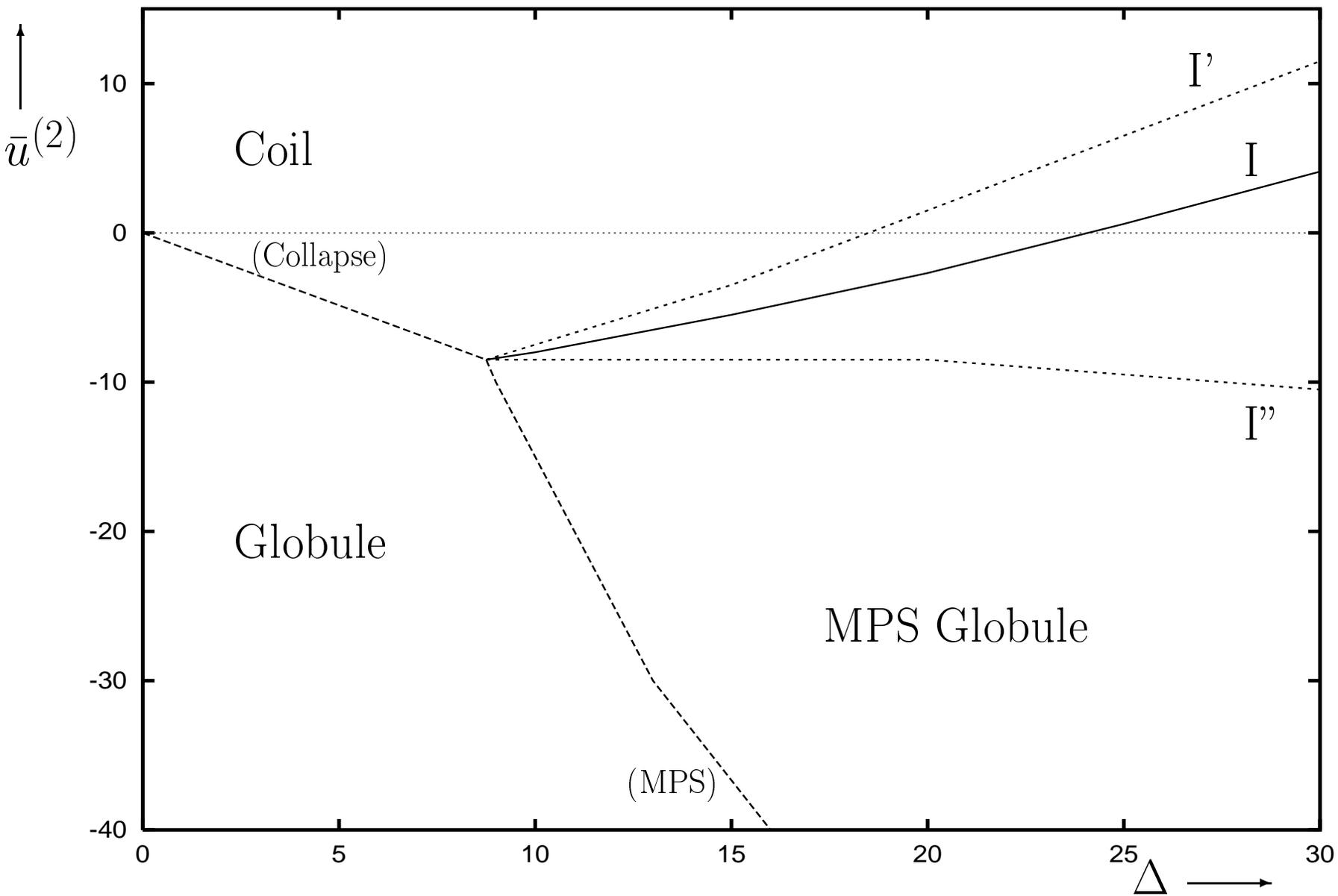
\begin{figure}
\caption{
The phase diagram of ``long block'' copolymer sequence '$(b_3 a_3)_5$'
in terms of the mean second virial coefficient, $\bar{u}^{(2)}$, and the
amphiphilicity, $\Delta$. Since the frustrated phases here are not accessible,
the collapse for large amphiphilicity proceeds through the discontinuous
transition (curve (I)) and it is accompanied by micro--phase separation.
Curves (I') and (I'') are spinodals.
}
\label{fig:lon30}
\end{figure}

Typical phase diagrams in terms of the mean second virial coefficient,
$\bar{u}^{(2)}$, and the amphiphilicity, $\Delta$, in Eq. (\ref{eq:u2})
for some ``random'' and periodic sequences are presented in
Figs. \ref{fig:ran30}-\ref{fig:lon30}.
In the region $\bar{u}^{(2)} > 0$ and for small values of amphiphilicity,
$\Delta < 5$, typical conformations of copolymers are akin to the
homopolymer extended coil.
By decreasing $\bar{u}^{(2)}$ to the negative region the chain undergoes the
continuous collapse transition, similarly to what we observed in
Sec. \ref{sec:si}.
The collapse transition is characterised by a rapid fall of the radius
of gyration, $R_g^2$, (\ref{eq:rg2}) and the change of the fractal
dimension, $\nu$ (see Tab. \ref{tab:si}). 

The collapse transition for larger values of amphiphilicity turns out to be
more complicated, and essentially dependent on the sequence. The globular
state for large values of $\Delta$ is different from the liquid--like
globule.
It is characterized by somewhat higher value of the radius of gyration
and extremely large value of the MPS order parameter (\ref{eq:mps}), thus we
call this state the MPS globule. The MPS globule is separated from the
liquid--like one by a weak continuous transition
(see Figs.~\ref{fig:ran30}-\ref{fig:lon30}).
In the case of long blocks (Fig.~\ref{fig:lon30}) the collapse
transition to the MPS globule becomes discontinuous (first--order--like).
The spinodals I' and I'' designate the region where two distinct states
corresponding to the coil and the MPS globule can be found.
The depths of the free energy minima become exactly 
equal on the transition curve I in Fig.~\ref{fig:lon30}.

However, for a wide class of sequences, for example for aperiodic sequences
in Figs. \ref{fig:ran30}, \ref{fig:ran60}, the phase diagram at large
amphiphilicity, $\Delta$,
is much more complicated. Starting from some value of $\Delta$ in some
intermediate region of $\bar{u}^{(2)}$ there appear additional solutions
corresponding to local minima of the free energy. The broad region
where this could take place is bounded by the curves I' and II'' in
Figs.~\ref{fig:ran30}, \ref{fig:ran60}.
With increasing $\Delta$ the number of such solutions grows quickly.
Significantly, in the region of the phase diagram, between curves I and II in
Figs. \ref{fig:ran30}, \ref{fig:ran60}, some of these possess the lowest
free energy value, thus being the thermodynamically stable state.
Since the number of such solutions is rather high even for short sequences
and their number grows quickly with the chain length, we do not attempt to draw
all their boundaries of (meta)stability. We shall call them collectively as
{\it frustrated phases}, explaining this terminology below.

Now let us compare the phase diagrams in Figs. \ref{fig:ran30} and
\ref{fig:ran60}, the latter corresponding to the sequence twice longer
than the for former. An interesting observation is that the region between
spinodals I' and II'', designating where the frustrated phases can exist,
expands dramatically with increasing chain length. The same concerns the
region of thermodynamically stable frustrated phases between curves I and II. 
More exactly, these regions expand downwards and to the left, so that the
position of curves I' and I change slightly with increasing $N$, whilst
curves II and II'' depend significantly on the size of the system.
For rather long chains we may expect that the regions of stability and
metastability of the frustrated phases will continue to expand downwards
and to the left, so that the lines II and II'' will become nearly vertical,
displacing the region of stable MPS globule. Probably, for most of long
heteropolymer chains the MPS globule does not exist as thermodynamically
stable state, becoming stable only for some special sequences.
Unfortunately, we can not proceed with numerical solution for much larger
system sizes, $N$, since the calculational expenses grow in $N$ as $N^3$
per iteration and also the total number of frustrated states becomes
huge for large system sizes.
This diversity and a special foliating structure of various branches leads
in the thermodynamic limit to what is known as a spin glass frozen phase 
\cite{Mezard-book} of random copolymers.

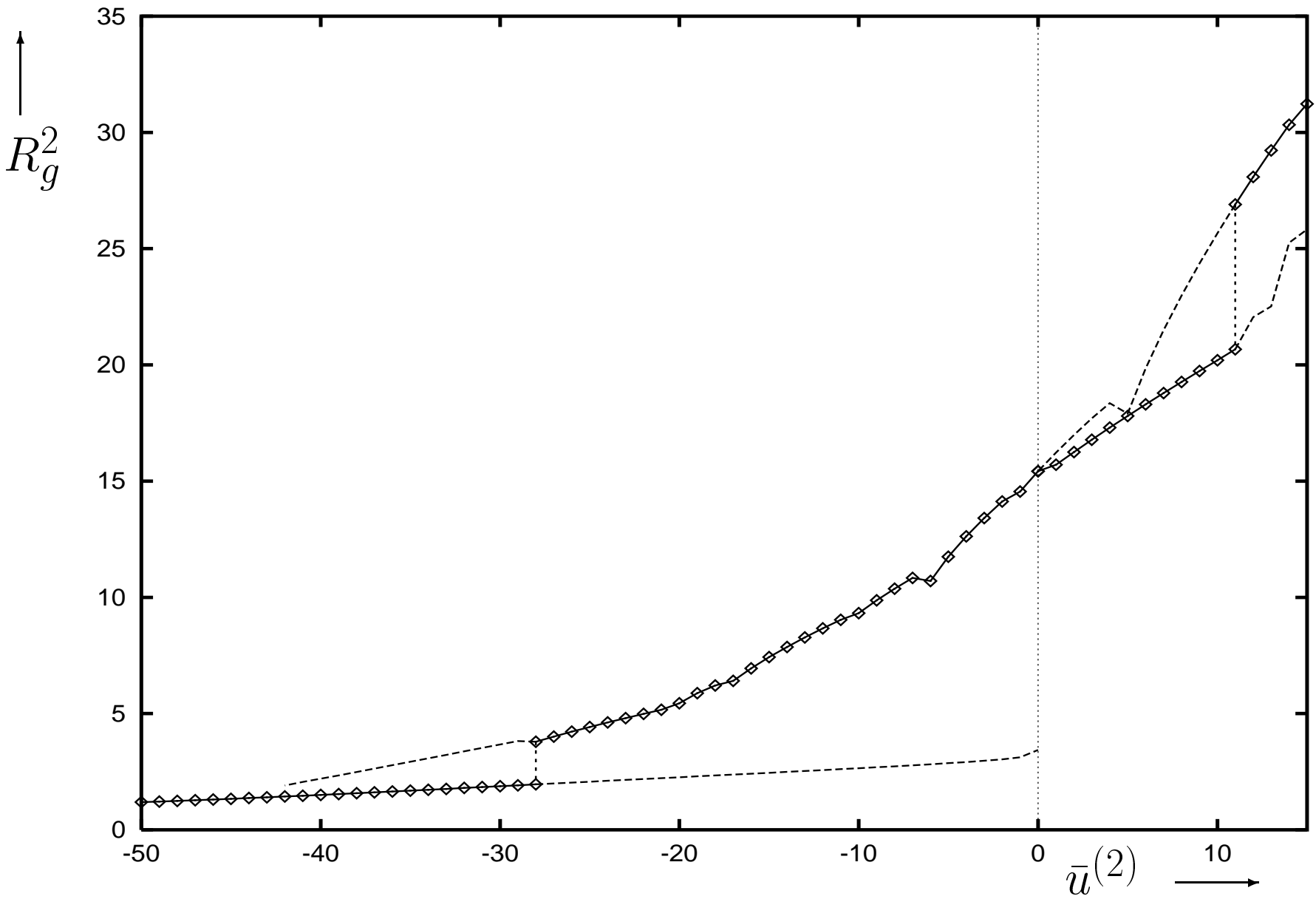
\begin{figure}
\caption{
Plot of the mean squared radius of gyration, $R_g^2$, versus the mean second
virial coefficient, $\bar{u}^{(2)}$, for '$(ab)_{30}$' copolymer.
Here and in Fig.~\ref{fig:mps} $\Delta = 30$, solid lines correspond
to values of observables in the main free energy minimum and dashed
lines --- in metastable minima.
}
\label{fig:rg2}
\end{figure}

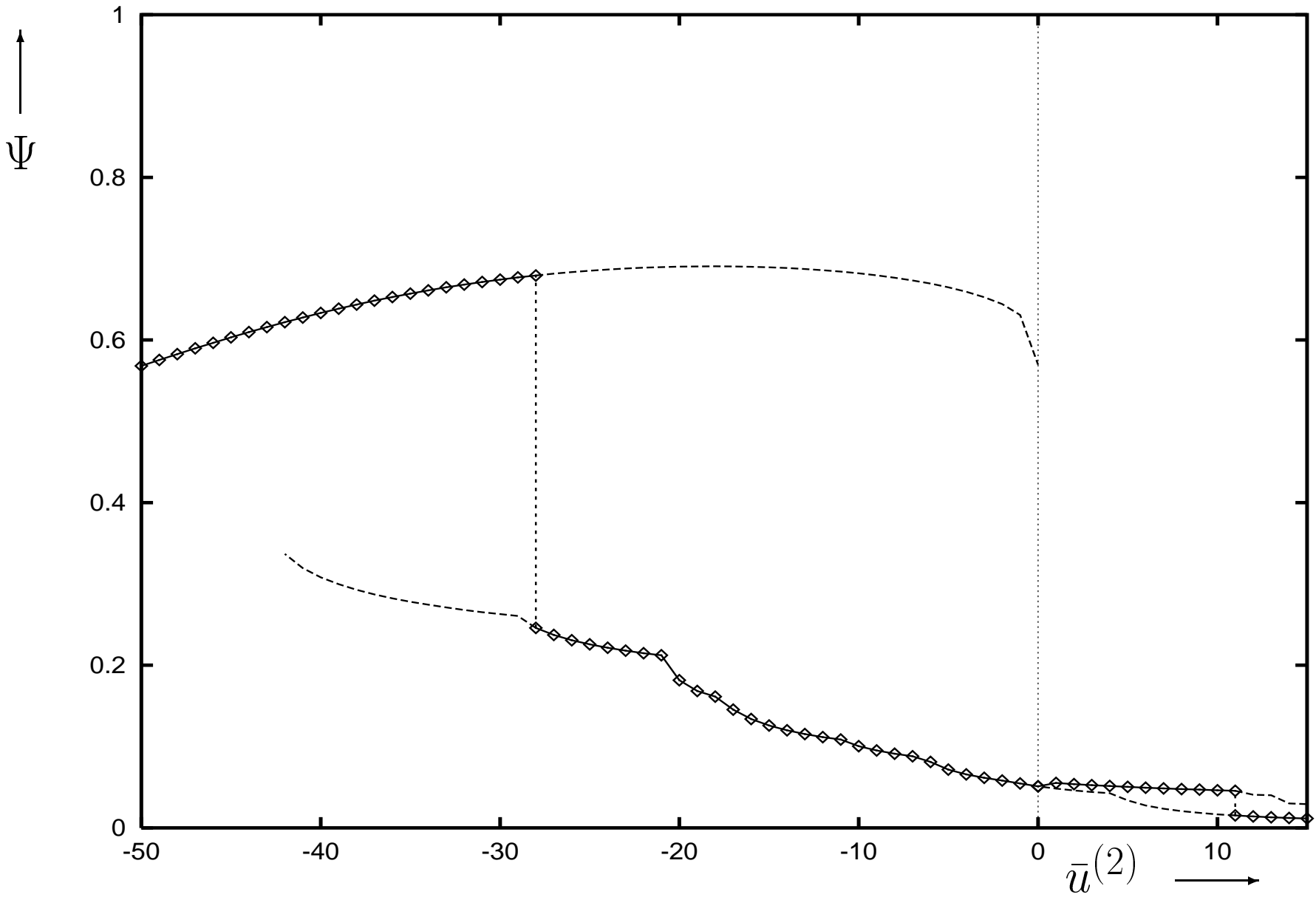
\begin{figure}
\caption{
Plot of the parameter of micro--phase separation, $\Psi$, versus the mean
second virial coefficient, $\bar{u}^{(2)}$, for '$(ab)_{30}$' copolymer.
See also caption to Fig.~\ref{fig:rg2} for more details.
}
\label{fig:mps}
\end{figure}

Let us consider the conformational structure of the frustrated states for
the copolymer consisting of repeating '$ab$' blocks. The phase diagram of
this sequence also exhibits the thermodynamically stable frustrated phases
\cite{GscArbCop} starting from approximately $N = 28$.
In Figs.~\ref{fig:rg2} and \ref{fig:mps} we present the dependence of the
observables in (\ref{eq:rg2}, \ref{eq:mps}) at a quasistatic change of
the mean second virial coefficient, $\bar{u}^{(2)}$, from the coil state to
the MPS globule and back. From these pictures one can see in the intermediate
region only a few (seven to be precise) of all possible frustrated phases.
The values of the radius of gyration and the MPS order parameter are
intermediate for these solutions, lying between those of the coil and
the MPS globule. In this sense, we can call them non--fully compacted
and misfolded states.

In the series of pictures in Fig.~\ref{fig:dmm} we present the matrix of
mean squared distances, $D_{mm'}$, for the copolymer, consisting of '$ab$'
blocks at some values of the mean second virial coefficient, $\bar{u}^{(2)}$.
The set of $D_{mm'}$ in the GSC method completely determines the
conformational structure of any, equilibrium or kinetic, state.
For positive $\bar{u}^{(2)}$ (see Fig. \ref{fig:dmm}a) the
mean squared distances possess the structure typical for the extended coil.
The elements of the matrix, $D_{mm'}$, increase
monotonically on moving away from its diagonals towards the distance of
half--ring along the chain.
Thus, the $D_{mm'}$ matrix may be approximated here by a monotonically
increasing function of the chain distance, $|m-m'|$.
Decreasing the mean second virial coefficient, $\bar{u}^{(2)}$, causes
the copolymer to pass through frustrated states, in
Figs. \ref{fig:dmm}b-\ref{fig:dmm}d,
finally reaching the MPS globule state (see Fig.~\ref{fig:dmm}e).
The characteristic feature of the $D_{mm'}$ matrix in a frustrated state is
that it possesses some number of monomer groups having smaller
distances between each other than between monomers from other groups.
Clearly, such a group represents a cluster of monomers, so that here
the copolymer chain forms a set of clusters (approximately 8 clusters in
Fig.~\ref{fig:dmm}b, 4 clusters in Fig.~\ref{fig:dmm}c and 2 clusters in
Fig.~\ref{fig:dmm}d), each consisting of monomers nearest along the chain.

The internal structure of each cluster is similar to the structure of the
final MPS globule. For the simplest copolymer sequence presented here,
the $D_{mm'}$ matrix away from the diagonal can be divided into
sub--matrices of the size $2\times 2$ (see top of the Fig.~\ref{fig:dmm}e),
each of which possesses approximately the same structure:
small values in the upper left corner correspond to the
mean squared distances between hydrophobic monomers; large values in the
lower right corner, which are the mean squared distances between hydrophilic
monomers; and off-diagonal elements are nearly equal and correspond to the
distances between different species. Thus, for the MPS globule, such
structure of the $D_{mm'}$ matrix reflects the structure of the
two--body interaction matrix, $u^{(2)}_{mm'}$. Obviously, the higher
correlation between these two matrices is manifested in the higher value of
the MPS order parameter. For another, more complicated, sequences the
$D_{mm'}$ matrix in the MPS globular state has a different structure, but still
it resembles in some way the interaction matrix, $u^{(2)}_{mm'}$ away from
the diagonal, being distorted by the spring interactions in the elements
close to the diagonal. Note that for the liquid--like globule the pattern
of the $D_{mm'}$ matrix looks sufficiently trivial. Namely, it has narrow
band of dark color along the main diagonal with its intensity quickly
decreasing, as one can see from Fig.~\ref{fig:dmm}f.

From Fig.~\ref{fig:dmm} one can see that the interaction symmetries imposed
by the EFEF (\ref{eq:H}) are spontaneously broken in the region of the
phase diagram corresponding to the frustrated phases. Obviously, in our
case the number of possible ways to break the symmetry is enormously huge
and moreover it grows exponentially with increasing system size.
Despite the kinematic symmetries are not present for arbitrary sequences,
the structure of the phase diagram (see Figs.~\ref{fig:ran30}, \ref{fig:ran60})
and behaviour of main observables remain very similar.
It is the particular structure of $D_{mm'}$ matrix, number and 
the shape of boundaries of frustrated phases
that are quite sensitive to the sequence.
The symmetry that may be broken in this case has a subtler meaning
and may be expressed in terms of the replica formalism \cite{Mezard-book}.
Consider, for example, two nearly identical blocks with nearly identical
surroundings. In the coil and the MPS globular states one should expect the
conformational matrices of these blocks to be nearly equal to each other.
On the contrary, the small  difference in the interactions
in the region of frustrated phases may lead to a huge difference in the
conformations. Thus, it is by no means surprising that the replica symmetry
breaking in the case of periodic systems takes such an explicit
manifestation in the breaking of the block translational symmetry.

An important point here is that the frustrated phases become dominant 
in an intermediate region of the phase diagram not due to a low mean energy,
but mostly due to a higher entropy.
The MPS globule is entropically unfavourable there because the
overall shrinking force is insufficiently strong.
Also, in the regime of nearly compensating repulsion and attraction between
monomers it is more preferable to achieve phase separation on a smaller,
than the globular, scale by forming clusters.

\section{Folding Kinetics}
\label{sec:kin}

Here we shall consider the time evolution of the conformational
state of a copolymer away from the initial equilibrium 
after it has been subjected to an instantaneous
temperature jump that causes the two--body interaction parameters in
Eq. (\ref{eq:u2}), $\bar{u}^{(2)}$ and $\Delta$, to change. The composition
of copolymer $\{\sigma\}$ and the rest of parameters remain the same and
do not change with time. We are interested in quenches
from the homopolymer coil, where all monomers are equally 
hydrophilic ($\bar{u}^{(2)} >0$ and $\Delta=0$), to the region of parameters
corresponding to the MPS globular state, so that the `$a$' species became
strongly hydrophobic and the `$b$' species remained nearly neutral
($\bar{u}^{(2)} \ll 0$ and $\Delta \approx |\bar{u}^{(2)}|$).
Here we consider some binary copolymer systems, consisting of the same number
of '$a$' and '$b$' monomers and of the same size, $N = 50$, and quench,
$(\bar{u}^{(2)}=15,\ \Delta=0) \rightarrow (\bar{u}^{(2)}=-35,\ \Delta=30)$.

Let us first discuss the general features of the kinetics of the copolymer
folding. The time evolution of the mean squared radius of gyration,
$R_g^2 (t)$, the MPS order parameter, $\Psi (t)$,
and the instantaneous free energy, ${\cal A} (t)$,
is presented in Figs.~\ref{fig:set1}, \ref{fig:set2}.
Here lines A correspond to the homopolymer and serve for reference purpouses.
In the case of the homopolymer one can see that $R_g^2$ and ${\cal A}$
decrease monotonically to their equilibrium values corresponding to
the liquid--like globule state, while $\Psi$ remains identically zero.
As for the copolymer kinetics, the first observation consists in that
it proceeds much slower than for the homopolymer. For example, for the
simplest copolymer sequence, '$(ba)_{25}$', the total collapse time
\cite{Footnote1} is more than 3 times longer than that of the homopolymer,
other copolymer sequences collapsing even longer. The total collapse time
also seems to be quite sensitive to the copolymer sequence.
Since the micro--phase separation is one of the main factors in collapse
of copolymers, the MPS order parameter, $\Psi$, grows in kinetics most of the
time for all considered sequences, though rather nonmonotonically.
Nevertheless, for some sequences it might
be negative during some time in kinetics (see e.g. sequences E and F in
Fig.~\ref{fig:set2}b), something we never observed at equilibrium.

The evolution of the instantaneous free energy, ${\cal A} (t)$, depicted in
Figs.~\ref{fig:set1}c, \ref{fig:set2}c, is most unusual.
Typically it proceeds through multiple accelerations and decelerations.
The flat regions of a staircase--like function correspond
to temporary kinetic arrest of the system in transient nonequilibrium 
conformations, i.e.
to transient trappings of various members of the ensemble in their local
shallow energy minima. Since such minima are encountered at
different moments in time for different members of the ensemble,
their influence on the overall time evolution of averaged observables
is manifested in a smooth characteristic slowing down.
Note here that the number of steps in kinetics process hardly can be
guessed from the given primary sequence. Typically, the kinetics for
sequences with smaller number of blocks proceeds through smaller number
of steps. For example, folding of the periodic sequence consisting
of long blocks, '$(b_5 a_5)_5$ (see line E in Fig.~\ref{fig:set2}c),
 proceeds through only one, though rather
long, kinetically arrested step. However, for the aperiodic sequence of
long blocks (line F in Fig.~\ref{fig:set2}c) we can see at least
six such steps, the third one being the longest in time.

Now let us consider the nonequilibrium conformations, given by the matrix of
the mean squared distances between monomers, $D_{mm'}$. In Fig.~\ref{fig:kind}
we exhibit the $D_{mm'}$ matrix for '$(ba)_{25}$' sequence at different
times in the folding kinetics. We can see that kinetic process proceeds
through formation of locally collapsed and phase--separated clusters.
The initial conformation is similar to Fig.~\ref{fig:dmm}a, then at
time $t = 3.6$ we can see approximately 10 clusters (Fig.~\ref{fig:kind}a),
at time $t = 5.75$ --- 6 clusters (Fig.~\ref{fig:kind}b) and
at $t = 8.6$ --- 4 clusters (Fig.~\ref{fig:kind}c).
During later evolution these four clusters approach each other since,
as one can see from Fig.~\ref{fig:kind}d, the inter--cluster distances are much
smaller than in Fig.~\ref{fig:kind}c. Finally, the clusters unify
forming the MPS globule, the $D_{mm'}$ matrix of which is quite similar to
that presented in Fig.~\ref{fig:dmm}e.
We should note here also that these nonequilibrium states do not possess the
translational block symmetry, present for the system in the EFEF and the
initial conditions. After some time in kinetics, $t\approx 1$
this symmetry breaks down, and restores only during final kinetic stages
at time moment $t\approx 15$.

However, the formal structure of the GSC kinetic equations (\ref{eq:kinmain})
is such that they yield a symmetric solution at any moment in time if
one proceeds from the symmetric initial condition.
What we observe here is a spontaneous symmetry breaking in kinetics.
Namely, at some moment in time the symmetric solution of the kinetic equations 
becomes unstable with respect to perturbations, whether of the initial 
condition, or of the interaction matrix, $u^{(2)}_{nn'}$.
In the exact theory there are fluctuations that can transform between
different spontaneously broken states in kinetics, thus destroying the
unstable symmetric state.
To describe such phenomena strictly in the framework of the Gaussian method,
which presents an improved, but still a mean--field type of theory, we should
include explicitly an infinitesimal symmetry breaking term $\varepsilon_{nn'}$
to the two--body interaction matrix, $u^{(2)}_{mm'}$, and consider the limit
of vanishing perturbation in the solution.
In fact, in numerical integration such a regularisation procedure
is not even necessary, since there is always an intrinsic perturbation 
due to computer round off and numerical integration errors. 
Thus, if the symmetry is favourable to be kinetically broken somewhere, 
numerically one obtains some spontaneously broken solutions there,
rather than the unstable symmetric solution, unless the symmetry
conditions have been imposed by hand. 
In fact, adding a rather small perturbation matrix, $\varepsilon_{nn'}$,
changes the behaviour of the global observables, such as
$R_g^2$, $\Psi$, ${\cal A}$ and characteristic kinetic times
rather insignificantly. However, what can be changed by
including of a perturbation is the centers along the chain
around which clusters form and grow.

Finally, let us discuss how the folding kinetics depends on the sequence.
In series of pictures in Fig.~\ref{fig:set1} we present kinetic processes
for the simplest copolymer consisting of '$ab$' blocks (sequence B),
and also for some sequences obtained by certain modifications in it.
In the sequence denoted by C in Fig.~\ref{fig:set1} we have
replaced ten short blocks by four of longer size.
In the sequence D we have inserted only two hydrophobic and
two hydrophilic fragments into the sequence, i.e. we have made only two
permutations. For the former sequence the
total kinetic time became approximately 1.7 times longer than that for the B
sequence. The final values of the free energy for both sequences B and C
are nearly the same, whilst the micro--phase separation is somewhat better
for the modified sequence C, due to longer blocks. Thus, for that sequence
the kinetic properties do not deteriorate very much, except that kinetic
process takes essentially longer.
However, the kinetic foldability of the D sequence is much more poor than
for B and C sequences. In fact, here not only the kinetic process takes
much longer, but the final state is different from the MPS globule.
As one can see from Fig.~\ref{fig:kinbad} the final kinetic state for
the D sequence consists of two clusters. These clusters are connected by
two links, formed by nearly neutral fragments, '$b_2$'. Further collapse
of this conformation is unfavourable due to the entropy and partial
screening of hydrophobic monomers by nearly neutral species '$b$'.
Size of such misfolded state is larger and the MPS order parameter
is approximately twice smaller than what we could expect for the MPS
globule.
As we have already seen kinetics of sequences consisting of longer
blocks is also quite sensitive to the sequence. Some modifications of the
periodic sequence E in Fig.~\ref{fig:set2} to form aperiodic long blocks
sequence, F, result in somewhat longer, but much more complicated,
kinetic processes.

\section{Conclusion}
\label{sec:con}

In this paper we applied the extended GSC method to studying the equilibrium
and kinetic phenomena of some particular amphiphilic heteropolymer sequences.
First, we have revised the Gaussian theory to include the self--interaction
term (\ref{eq:Esi}).
Thus, the two--body term in our formalism describes the
effective interactions between monomers, the self--interaction term is
responsible for stability of the globular state, and the three--body
interactions between distinct monomers provide the correct scaling
of the size of the globule.
The structure of the homopolymer globule is now in better agreement with
the one obtained from other methods.
We have found that the kinetic laws of the ring homopolymer
do not change significantly due to introduction of the self--interaction
term, except for the final relaxation time, which
without hydrodynamic interaction
scales with the degree of polymerization as, $\tau_f \sim N^{4/3}$.

We have obtained typical phase diagrams for heteropolymers consisting of
short and long blocks. For sequences with rather frustrated interactions
along the chain, apart from the coil, the liquid--like and the micro--phase
separated globular states, we have discovered that
in a wide intermediate region of the phase diagram there may be a large
number of frustrated partially misfolded states.
Some of such states, the number of which grows exponentially with the chain
length, become the dominant thermodynamic state in rather narrow domains.
We may conclude that the transition to these states for large systems
corresponds to a glassy freezing transition.

For large amphiphilicity parameter collapse from the extended coil state
to the MPS globule for a wide class of sequences proceeds at equilibrium
through these frustrated phases.
A typical copolymer conformation here is a set of micro--phase separated
clusters. On approaching the MPS globule quasistatically the number of
clusters decreases, reaching only one cluster in the final MPS globular state.
A typical cluster size is determined by some characteristic range where
the micro--phase separation may occur.

We have discovered that the region of meta-- and thermodynamic stability of
the frustrated phases expands with system size displacing the region of the
MPS globule. It is quite likely that the thermodynamically stable MPS
globule for heteropolymers of several hundreds of monomers is possible only
for a narrow class of some special sequences. Also it is likely that
inclusion of other interactions, such as electrostatic, and thus
modification of the two--body interaction matrix in Eq.~(\ref{eq:u2}),
may additionally stabilise the MPS globule.

We find that the kinetics of folding from the coil to the MPS globule
for copolymers takes much longer than for the homopolymer
since it is strongly affected by
the presence of the transient frustrated states along the kinetic pathway.
This leads to a complicated kinetic process consisting of multiple
steps with pronounced slowing down and then acceleration in the folding rate.
We also present here some preliminary studies on how the folding kinetics
depends on the primary heteropolymer sequence.
Typically, the kinetics for copolymers consisting of long blocks
proceeds in a smaller number of steps, but not necessarily faster,
than kinetics for short block copolymers.
For the latter we have seen that even small modifications of the
sequence may change crucially the overall kinetic behaviour and even
the final kinetic state itself.

There is an interesting arena here for classification of sequences
and conformational states. Moving in this direction would allow
to develop better models for proteins, which sequences have been
specially optimised by the biological evolution.

Finally, we hope that the way is now open for constructing the non--Gaussian
extension to the self--consistent method. 
This is difficult since the closure relations for higher--order
correlation functions are nontrivial for polymers.
When that is accomplished one could speak of real predictive accuracy
of the method.

\acknowledgments

The authors acknowledge interesting discussions with
Professor A.Yu.~Grosberg, Professor K.~Kawasaki, Professor G.~Parisi,
Dr~D.A.~Tikhonov, Dr~R.V.~Polozov %, Dr~R.~Reilly, Dr~T.~Kechadi
and our colleagues Professor K.A.~Dawson,  Dr~A.V.~Gorelov and A.~Moskalenko.
The authors acknowledge the support of the Centre for High Performance
Computing Applications, University College Dublin, Ireland.

\newpage

%%%%%%%%%%%%%%%%%%%%%%%%%%   Tables  %%%%%%%%%%%%%%%%%%%%%%%%%%%%%%%%%

\begin{table}
\caption{Values of the mean squared radius of gyration, $R_g^2$, for the
globular state at equilibrium, $u^{(2)} = -25$ and $u^{(3)} = 10$, and
characteristic collapse times, $\tau_t$ and $\tau_f$ after the quench
$u^{(2)} = 15 \,\rightarrow\,-25$ for different lengths of polymer
ring, $N$. Data for $c_3=3$ and $c_3=0$ correspond to the prescription with
and without self--interaction term (\ref{eq:Esi}). The last column contains
the effective exponent of the appropriate quantity in terms of the degree
of polymerization, i.e. $A \sim N^{\gamma}$.}
\label{tab:si}
\begin{tabular}{lcccccccc}
\hline
$N$             &   30  &   50  &   70  &  100  &  150  &  200  &  300  &
       exponent           \\
\hline
$R_g^2(c_3=0)$  &  1.19 & 1.91  & 2.59  & 3.53  &  4.96 & 6.25  &  8.52 &
$2 \nu = 0.86\pm 0.03$    \\
$R_g^2(c_3=3)$  &  0.91 & 1.32  & 1.68  & 2.17  &  2.90 & 3.55  &  4.72 &
$2 \nu = 0.69\pm 0.01$    \\
\hline
$\tau_t(c_3=0)$ &  3.9  & 10.5  & 20.3  & 40.8  &  90.7 & 160.3 &  357  &
$\gamma_t = 1.96\pm 0.01$ \\
$\tau_t(c_3=3)$ &  3.4  &  8.4  & 14.8  & 27.3  &  53.3 &  89.1 &  179  &
$\gamma_t = 1.70\pm 0.03$ \\
\hline
$\tau_f(c_3=0)$ &  1.18 &  2.97 &  5.49 & 10.2  &  20.6 &  33.0 & 65.7  &
$\gamma_f = 1.74\pm 0.03$ \\
$\tau_f(c_3=3)$ &  0.73 &  1.41 &  2.66 &  4.25 &  7.34 &  9.92 & 16.8  &
$\gamma_f = 1.27\pm 0.05$ \\
\hline
\end{tabular}
\end{table}

\vfill
\newpage

%%%%%%%%%%%%%%%%%%%%%%%%%%   Figures   %%%%%%%%%%%%%%%%%%%%%%%%%%%%%%%%

\begin{figure}
\caption{
Diagrams of the mean squared distances matrix, $D_{mm'}$ for '$(ab)_{30}$'
copolymer and amphiphilicity, $\Delta = 20$. Diagrams (a-e) correspond
respectively to the values of the mean second virial coefficient,
$u^{(2)} = 15$, $-11$, $-21$, $-30$ and $-40$.
Diagram (f) corresponds to the homopolymer globule, for which $u^{(2)} = -25$
and $\Delta = 0$. Indices $m$, $m'$ start counting from the upper
left corner. Each matrix element, $D_{mm'}$ is denoted by a quadratic
cell with varying degree of black colour, the darkest and the
lightest cells corresponding respectively to the smallest and to the largest
mean squared distances. The diagonal elements are not painted since
by definition, $D_{mm} = 0$.
}
\label{fig:dmm}
\end{figure}

\begin{figure}
\caption{
Plot of the mean squared radius of gyration, $R_g^2$ (Fig.~(a)), the MPS
order parameter, $\Psi$ (Fig.~(b)) and the instantaneous free energy,
${\cal A}$ (Fig.~(c)) versus time, $t$, during kinetics after the
quench from $\bar{u}^{(2)} = 15$, $\Delta = 0$ to
$\bar{u}^{(2)} = -35$, $\Delta = 30$ for copolymer sequences with
$N = 50$.  Lines A-D in the figures correspond respectively to the
following sequences: '$c_{50}$' (homopolymer), '$(ba)_{25}$',
'$a_2 b_3 a_3 b_2 a_3 b_2 a_2 b_3 (ba)_{15}$' and
'$(ba)_3 b_2 (ba)_9 a_2 (ba)_5 b_2 a_2 (ba)_4$'.
}
\label{fig:set1}
\end{figure}

\begin{figure}
\caption{
Plot of the mean squared radius of gyration, $R_g^2$ (Fig.~(a)), the MPS
order parameter, $\Psi$ (Fig.~(b)) and the instantaneous free energy,
${\cal A}$ (Fig.~(c)) versus time, $t$, during kinetics after the same
quench as in Fig.~\ref{fig:set1} for copolymer sequences with $N = 50$.
Lines A, E, F in the figures correspond respectively to the following
sequences: '$c_{50}$' (homopolymer), '$(b_5 a_5)_5$' and
'$b_6 a_4 b_5 a_5 b_4 a_6 b_3 a_7 b_7 a_3$'.
}
\label{fig:set2}
\end{figure}

\begin{figure}
\caption{
Diagrams of the mean squared distances matrix, $D_{mm'} (t)$ for
'$(ba)_{25}$' copolymer in kinetics after the same quench as in
Fig.~\ref{fig:set1}. Diagrams (a-d) correspond respectively to the
following moments in time: $t = 3.5$, $5.75$, $8.60$ and $13.0$.
See also caption to Fig.~\ref{fig:dmm} for more details.
}
\label{fig:kind}
\end{figure}

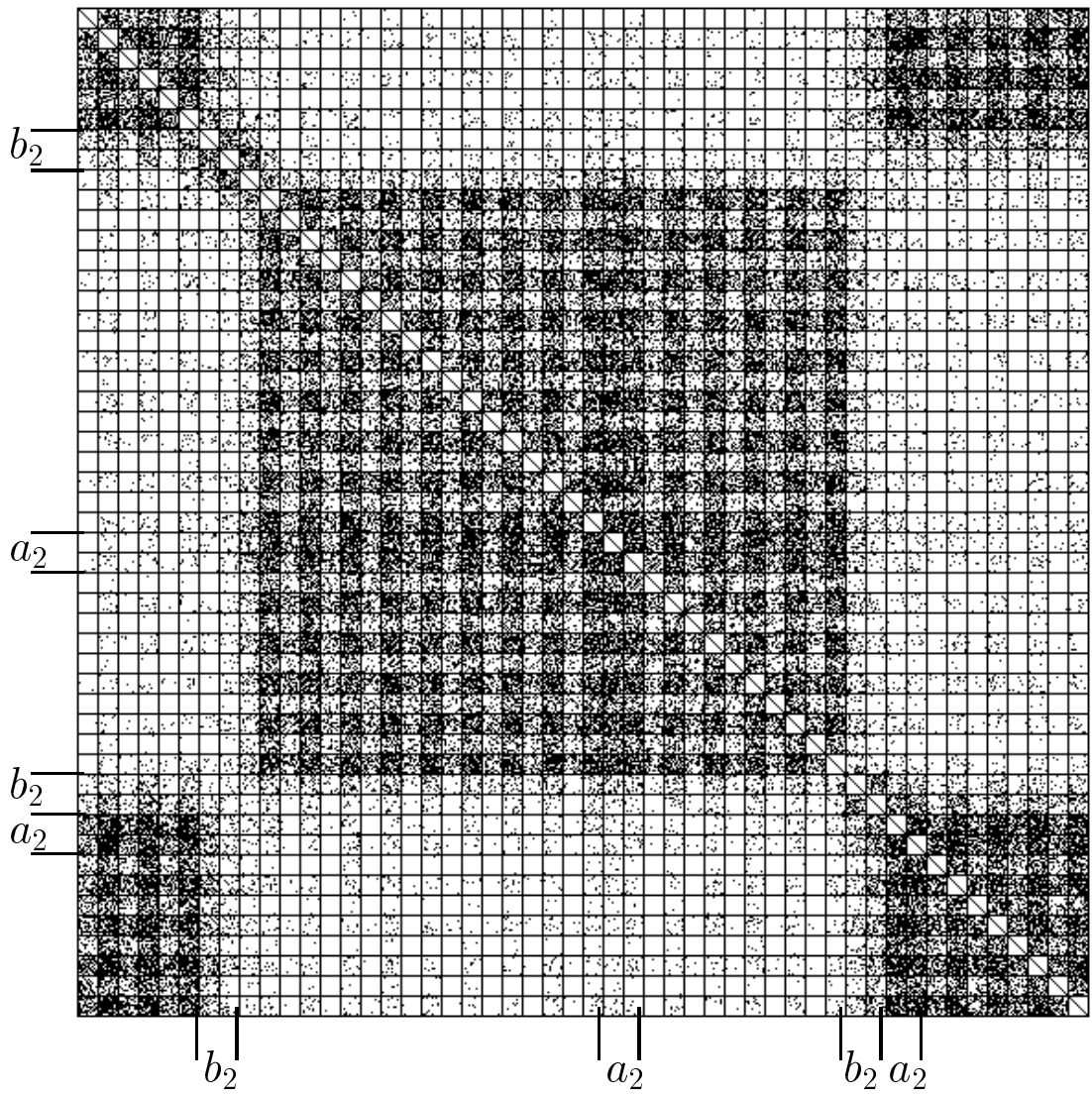
\begin{figure}
\caption{
The diagram of the mean squared distances matrix, $D_{mm'} (t)$ for
copolymer sequence '$(ba)_3 b_2 (ba)_9 a_2 (ba)_5 b_2 a_2 (ba)_4$'
at the time moment $t = 45$, so that this conformation is close to the
final equilibrium. Parameters of the quench are the same as in
Fig.~\ref{fig:set1}.
See also caption to Fig.~\ref{fig:dmm}, \ref{fig:kind} for more details.
}
\label{fig:kinbad}
\end{figure}


\begin{thebibliography}{00}

\bibitem{BlockCopolymers}
\BY{T.~Ohta, \atque K.~Kawasaki} \IN{Macromolecules}{19}{1986}{2621};
\BY{L.~Leibler} \IN{Macromolecules}{13}{1980}{1602};
\BY{G.~H.~Fredrickson, \atque E.~Helfand} \IN{J. Chem. Phys.}{87}{1987}{697};
\BY{T.~Hashimoto} \IN{Macromolecules}{20}{1987}{465};
\BY{T.~Garel, \atque H.~Orland} \IN{Europhys. Lett.}{6 (4)}{1988}{307};
\SAME{6}{1988}{597};
\BY{G.~H.~Fredrickson, S.~T.~Milner, \atque L.~Leibler}
 \IN{Macromolecules}{25}{1992}{6341}.

\bibitem{Blends}
\BY{M.~L.~Huggins} \IN{J. Chem. Phys}{9}{1941}{440};
\BY{P.~J.~Flory} \IN{J. Chem. Phys}{9}{1941}{660};
\TITLE{Polymer Blends}, edited by \BY{D.~R.~Paul, \atque S.~Neuman}
 (Academic Press, 1987);
\TITLE{Multicomponent Polymer Systems}, edited by
\BY{I~.S.~Miles, \atque S.~Rostami}
(Longman Scientific and Technical, Singapore, 1992).

\bibitem{GscKinet}
\BY{E.G.~Timoshenko, Yu.A.~Kuznetsov, \atque K.A.~Dawson}
\IN{J. Chem. Phys.}{102}{1995}{1816}.

\bibitem{GscHomKin}
\BY{Yu.A.~Kuznetsov, E.G.~Timoshenko, \atque K.A.~Dawson}
\IN{J. Chem. Phys.}{104}{1996}{3338}.

\bibitem{GscBlock}
\BY{E.G.~Timoshenko, Yu.A.~Kuznetsov, \atque K.A.~Dawson}
\IN{Phys. Rev.}{E 53}{1996}{3886}.

\bibitem{Allegra}
\BY{G.~Allegra, \atque F.~Ganazzoli} \IN{J.~Chem.~Phys.}{83}{1985}{397};
\BY{G.~Raos, \atque G.~Allegra} \IN{J. Chem. Phys.}{104}{1996}{1626}.

\bibitem{GscArbCop}
\BY{E.G.~Timoshenko, Yu.A.~Kuznetsov, \atque K.A.~Dawson}
submitted to \TITLE{Phys. Rev. E}.

\bibitem{Edwards}
\BY{M.~Doi, S.~F.~Edwards} \TITLE{The Theory of Polymer Dynamics}
 (Oxford Science, NY, 1989).

\bibitem{deGennes}
\BY{P.~G.~de Gennes} \TITLE{Scaling Concepts in Polymer Physics} (Cornell
 Univ. Press, NY, 1988).

\bibitem{Cloizeaux}
\BY{J.~des~Cloizeaux, \atque G.~Jannink} \TITLE{Polymers in Solution}
 (Clarendon Press, Oxford, 1990).

\bibitem{Footnote0}
In what follows we fix the units of temperature, size and time
by choosing $k_B T = 1$, $l = 1$ and $\zeta_b = 1$.
We account for the volume interactions up to three--body terms, i.e. we assume
that $u^{(J)}_{\{n\}} = 0$ for $J > 3$. Taking into consideration of the
self--interaction term (\ref{eq:Esi}), to prevent collapse of monomers for
any negative second virial coefficient, is sufficient if the third
virial coefficient is positive.
below we also fix the third virial coefficient, $u^{(3)} = 10$.

\bibitem{GrosbKhokh}
\BY{A.~Yu.~Grosberg, A.~R.~Khokhlov} \TITLE{Statistical Physics of
 Macromolecules} (AIP, NY, 1994).

\bibitem{CoplmMonte}
\BY{Yu.A.~Kuznetsov, E.G.~Timoshenko, \atque K.A.~Dawson}
\IN{J. Chem. Phys.}{102}{1995}{4807}.

\bibitem{Footnote1}
The total collapse time, $\tau_t$, defined as that time, when the squared
radius of gyration has passed through $99\%$ of its overall change:
$R_{g}^{2}(\tau_{t}) = 0.01\,R_{g}^{2}(0) + 0.99\,R_{g}^{2}(\infty)$.
The final relaxation, $\tau_f$, determined from the mean squared
radius of gyration $R_g^2 (t) = R_g^2 (\infty) + A_f\,e^{-t/\tau_f}$,
or equivalently the time--scale deduced from the first internal mode
${\cal F}_1(t)$ (see Ref. \cite{GscHomKin} for more details).

\bibitem{TorNew}
\BY{Yu.A.~Kuznetsov, E.G.~Timoshenko}
\IN{J. Chem. Phys.}{111}{1999}{3744}.

\bibitem{NumerRecip}
\BY{W.~H.~Press, S.~A.~Teukolsky, W.~T.~Vetterling, \atque B.~P.~Flannery}
\TITLE{Numerical Recipes in C} (Cambridge University Press, 1992).

\bibitem{Mezard-book}
\BY{M.~Mezard, G.~Parisi, \atque M.~Virasoro} \TITLE{Spin glass theory
 and beyond} (World Scientific, Singapore, 1987).

\end{thebibliography}
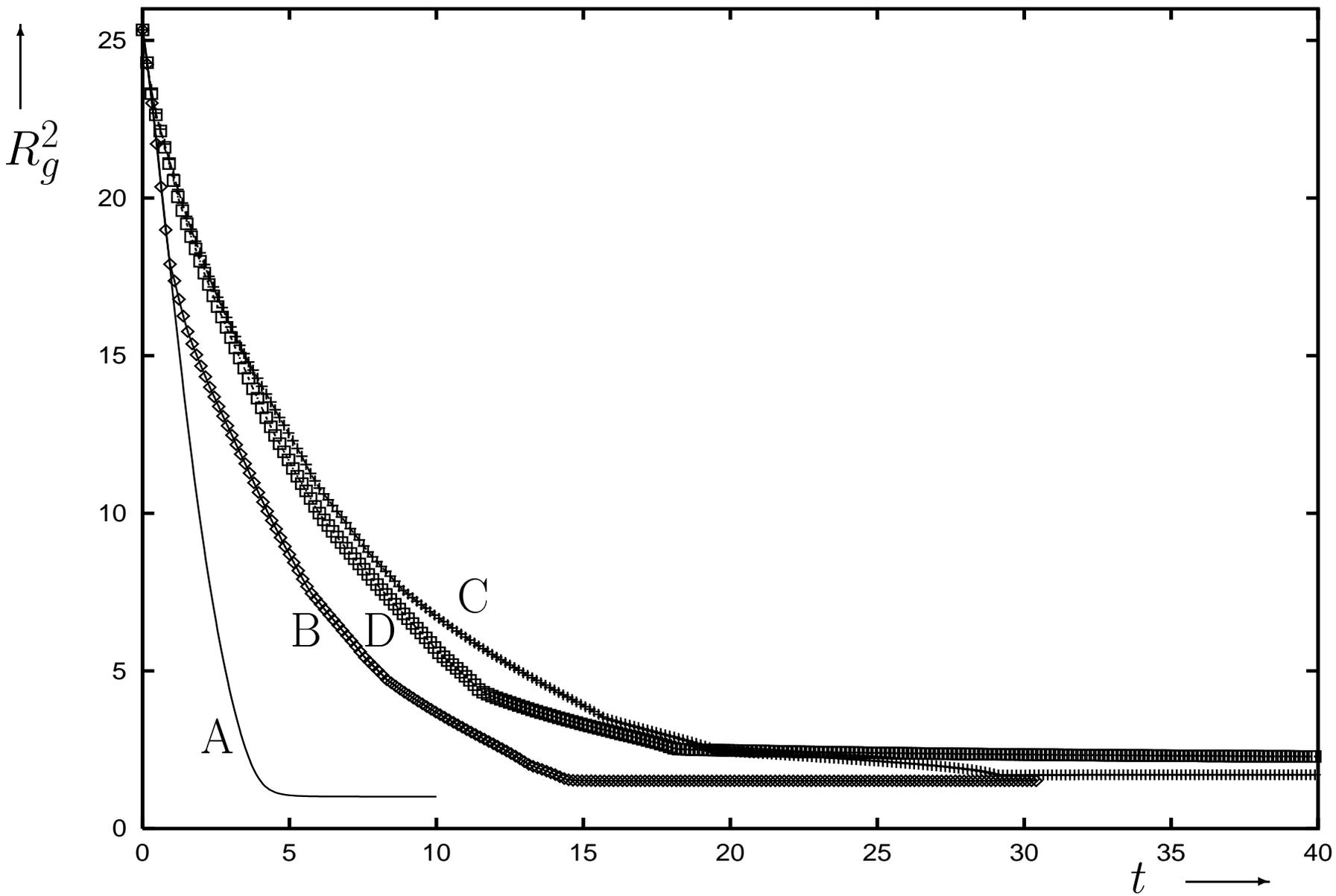
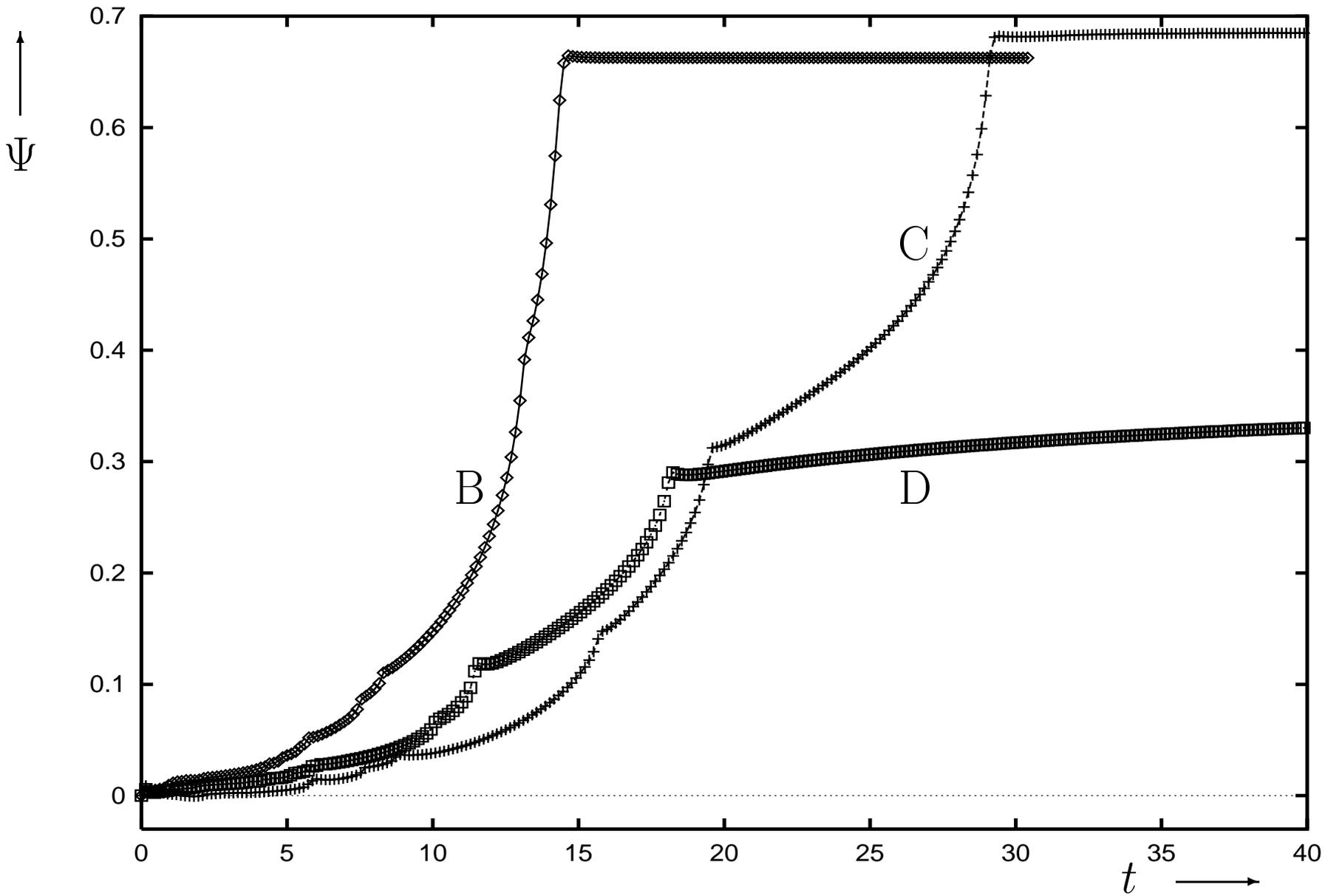
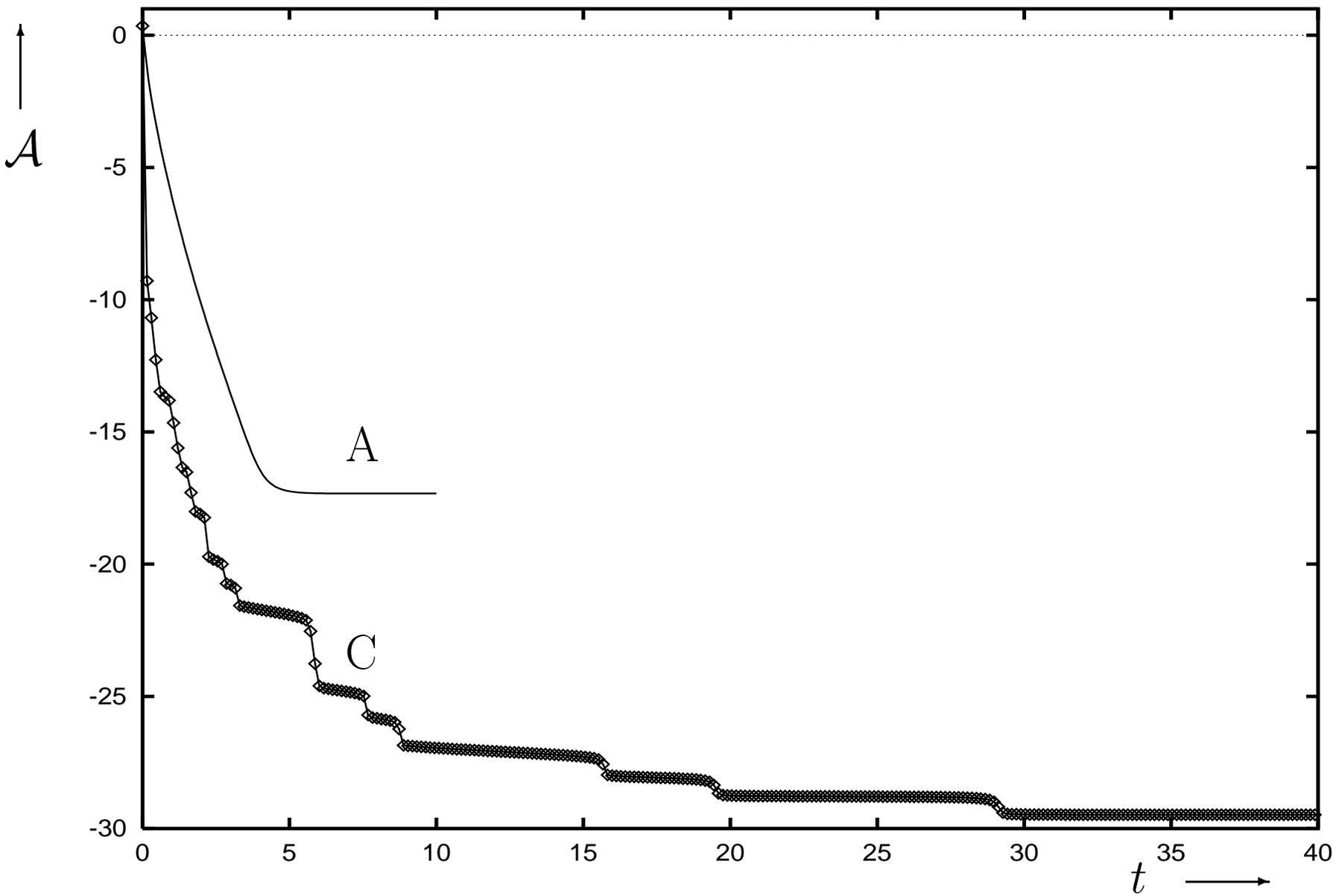
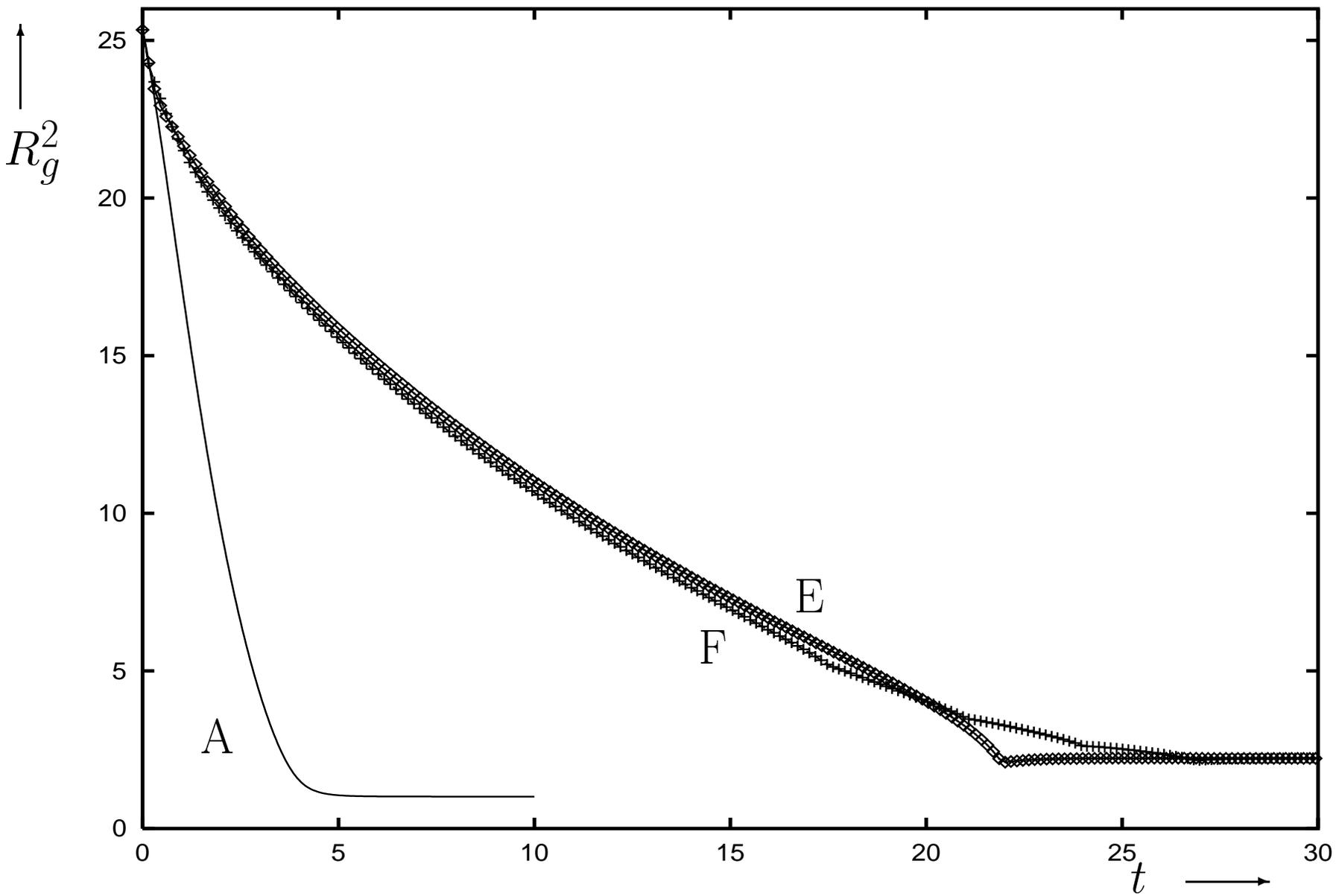
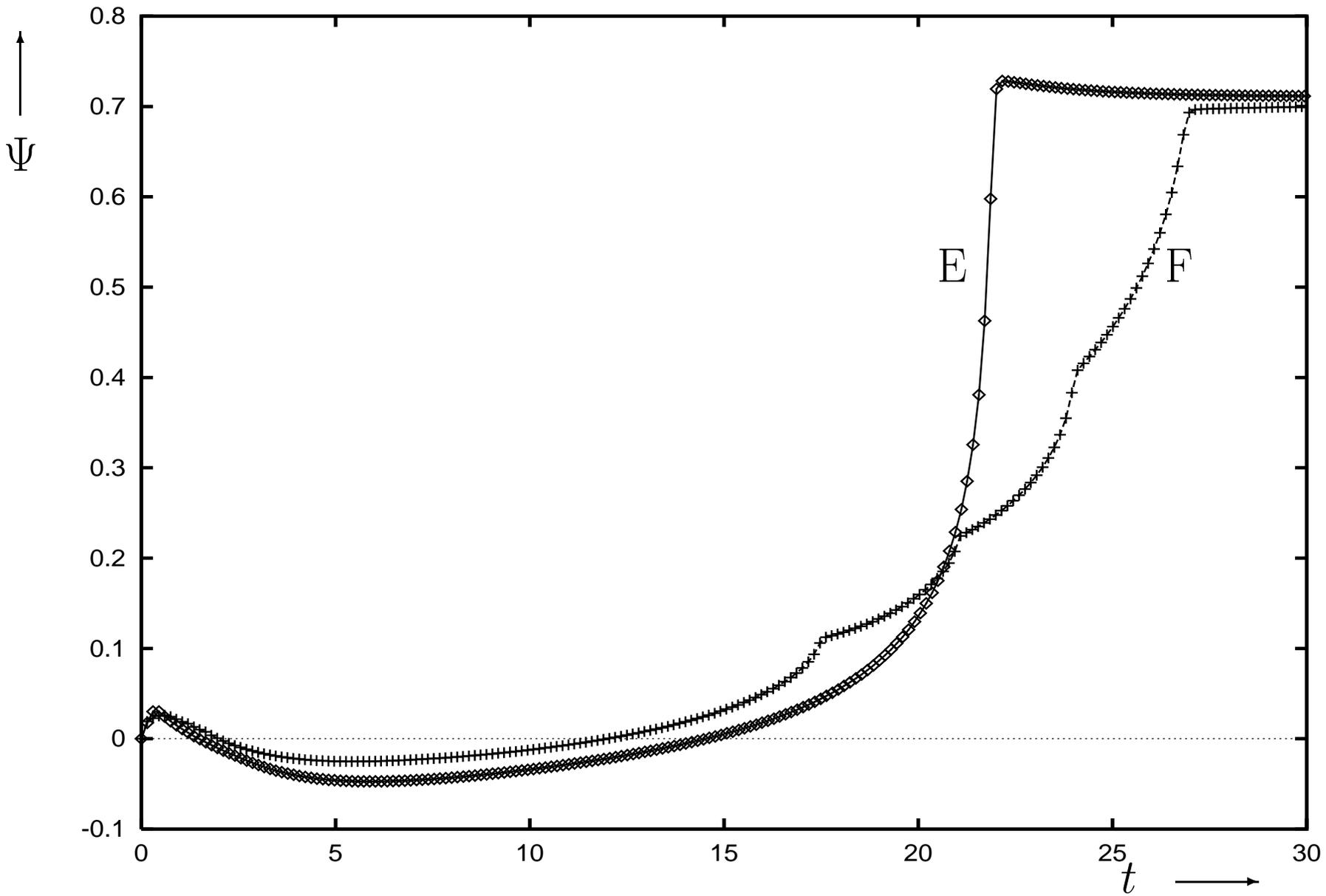
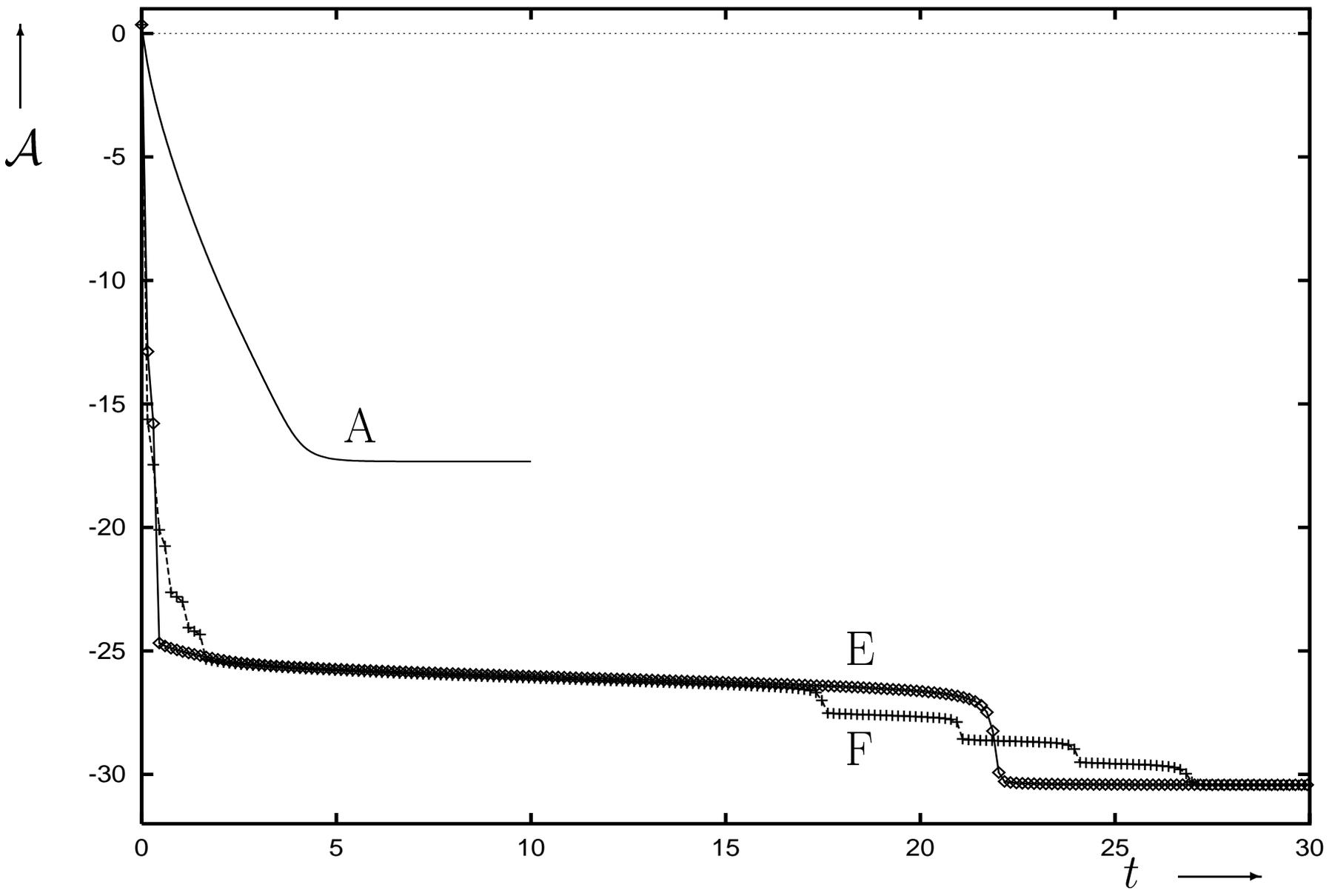
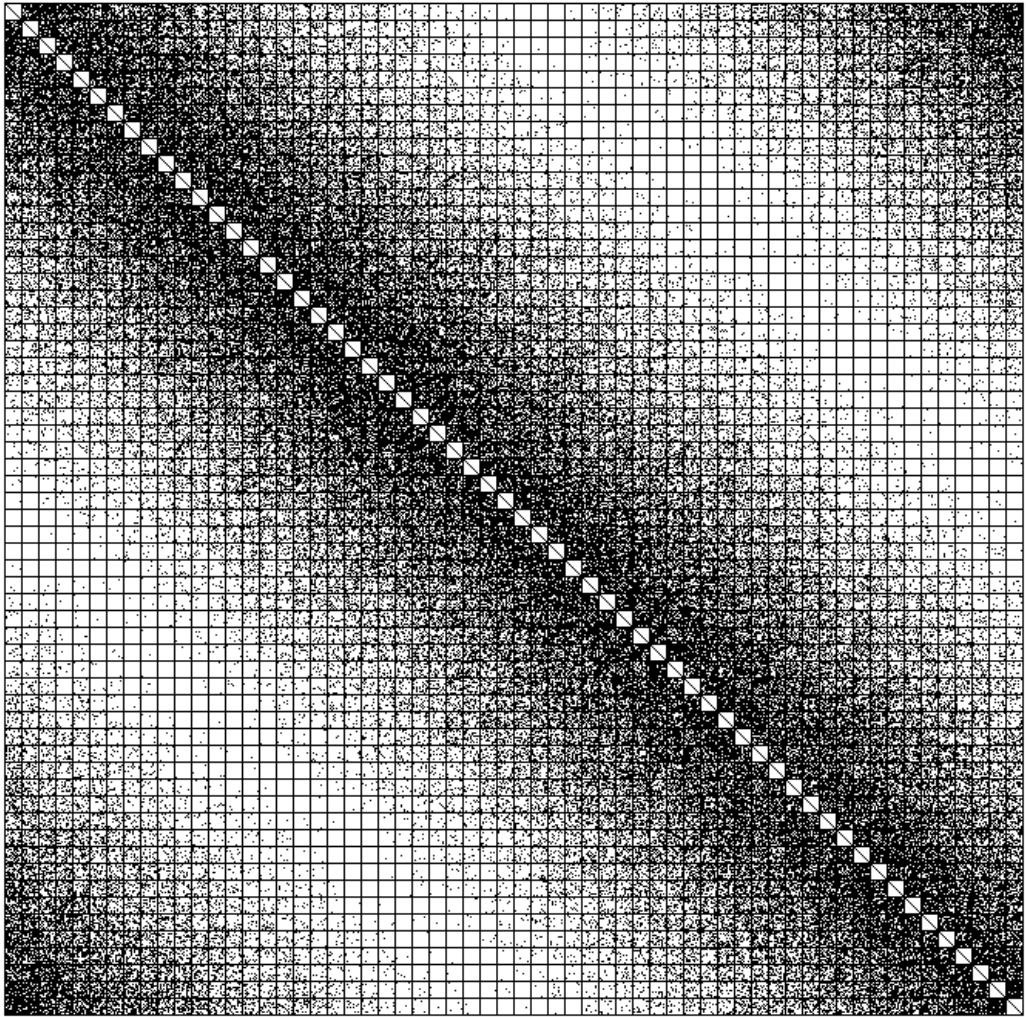
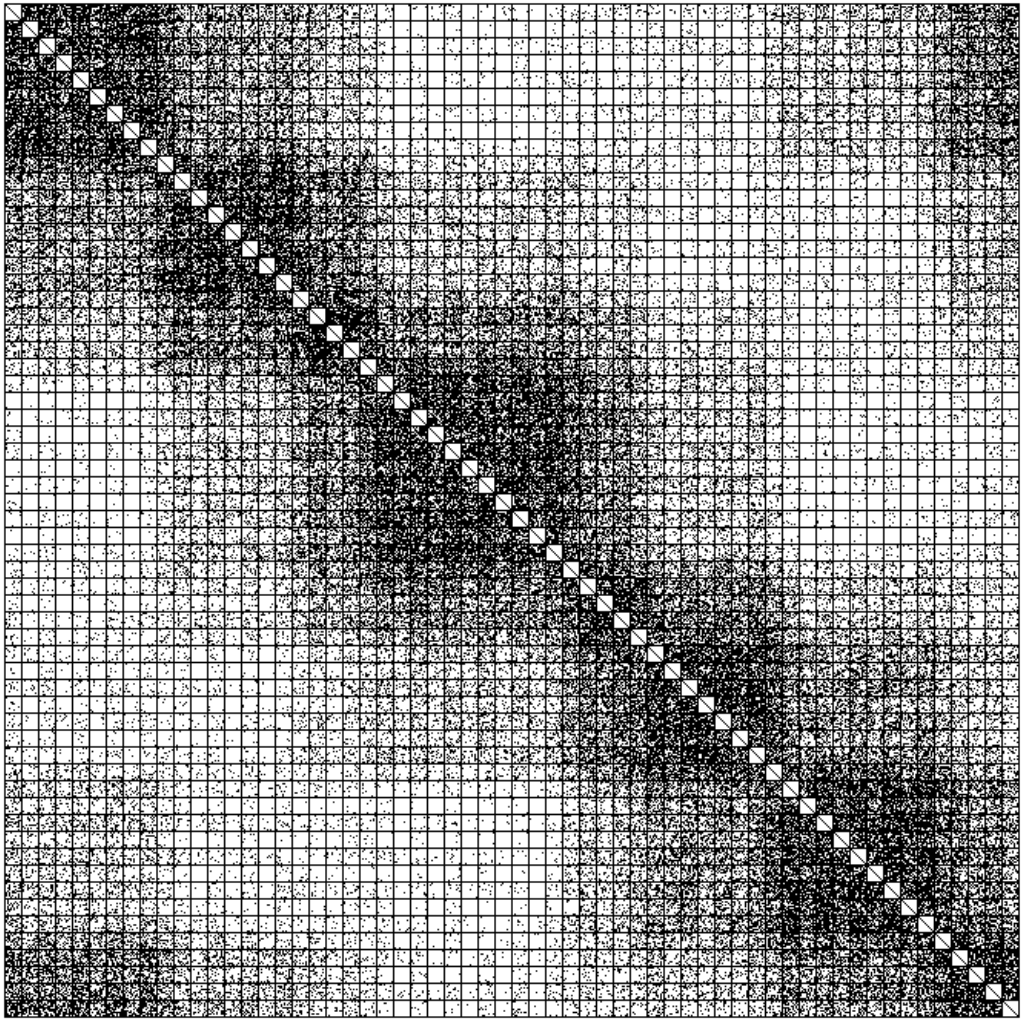
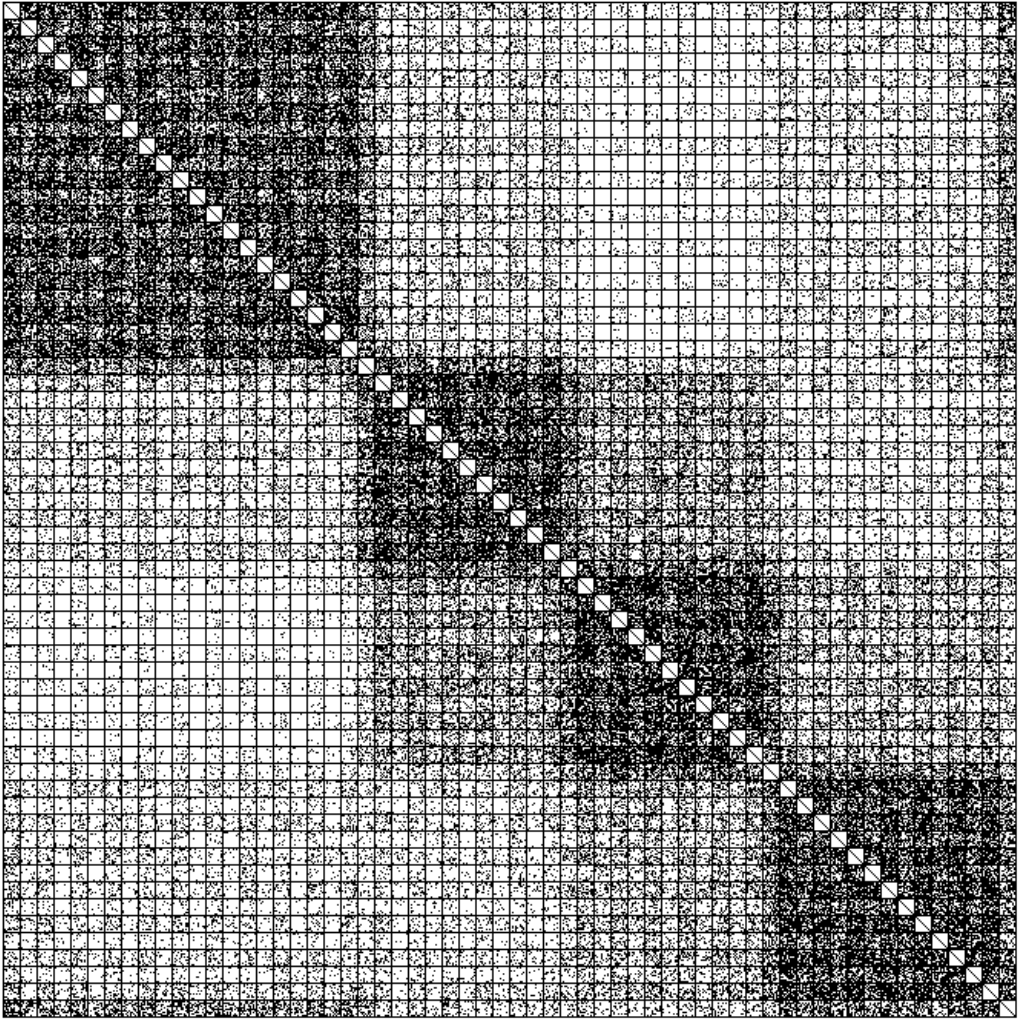
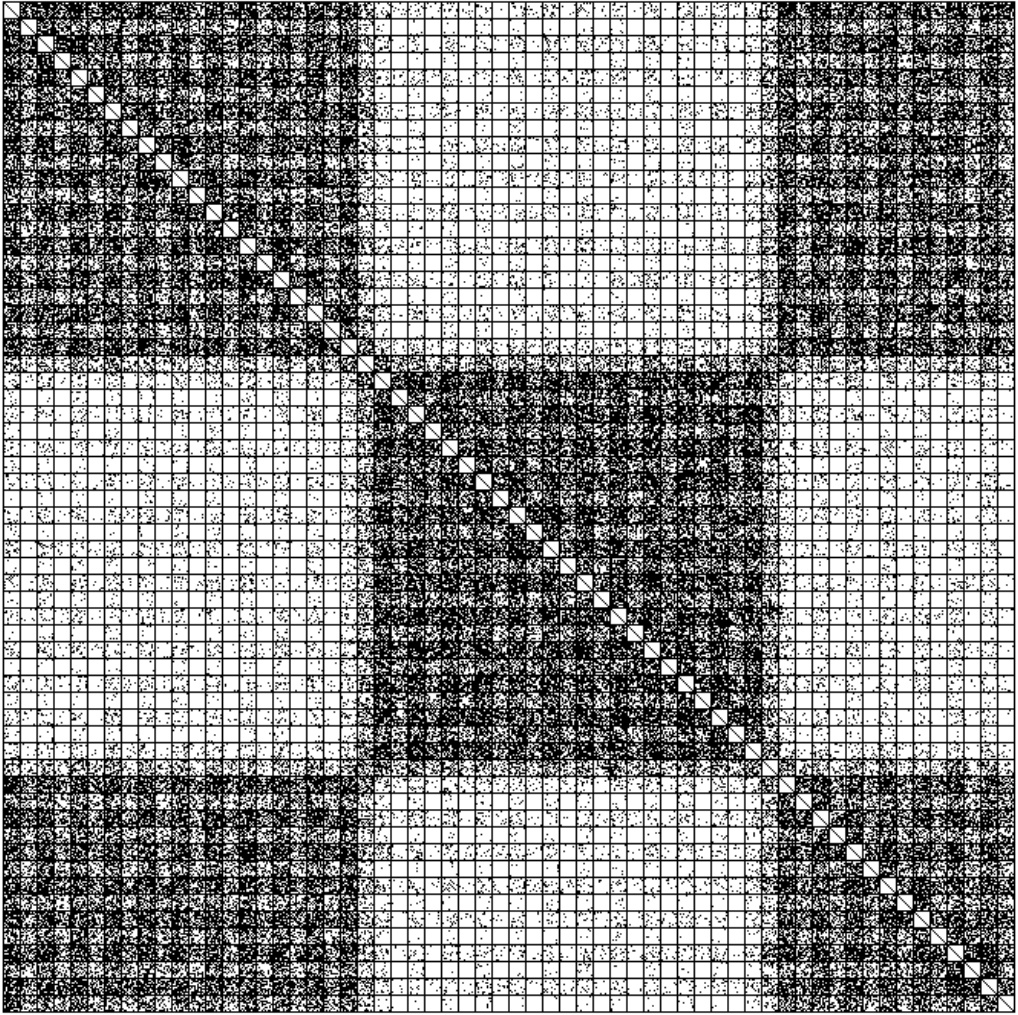
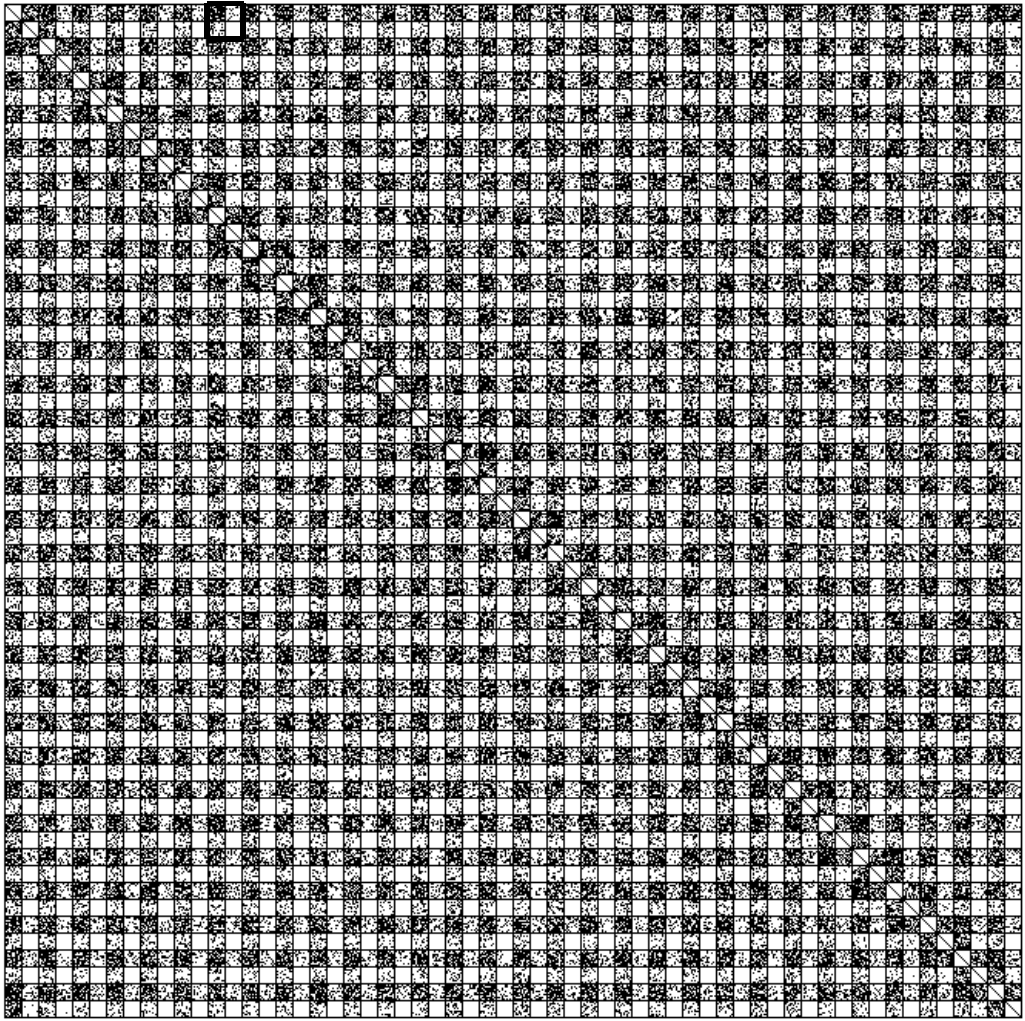
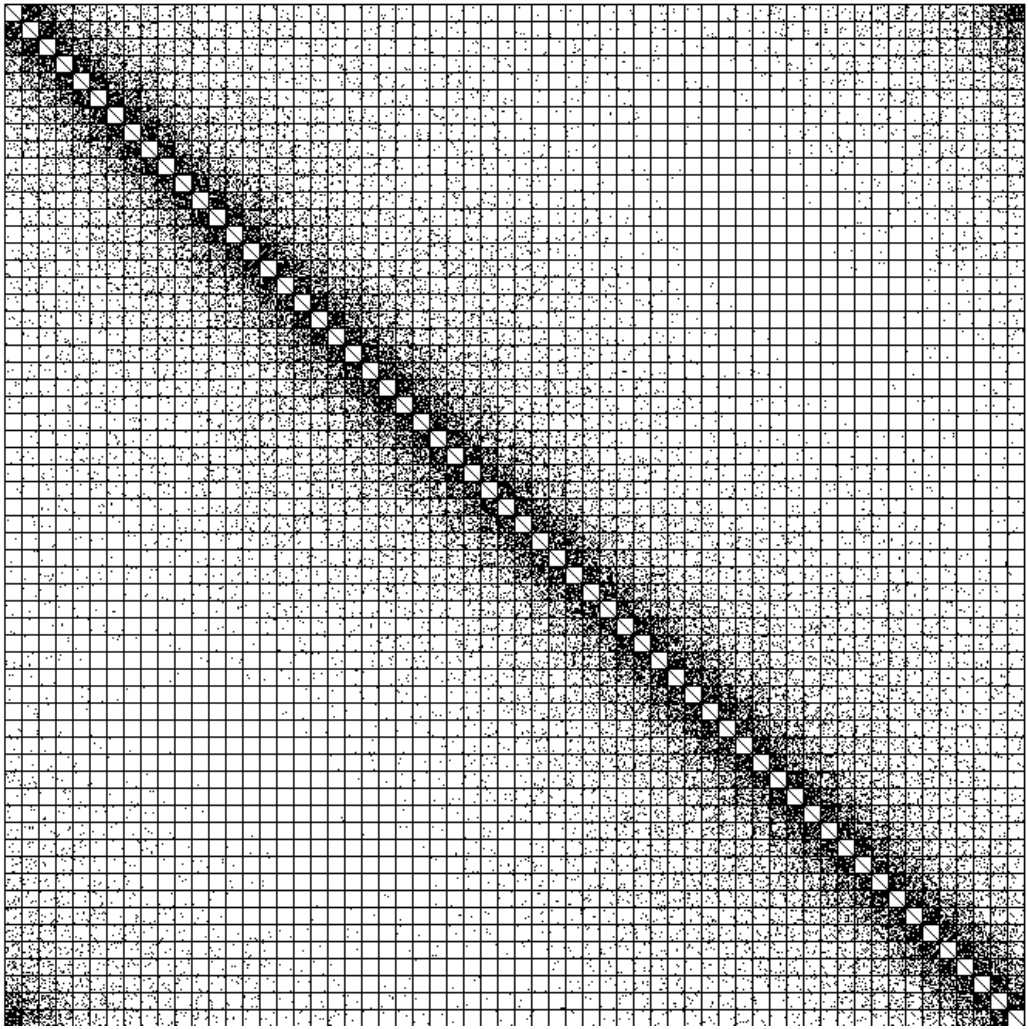
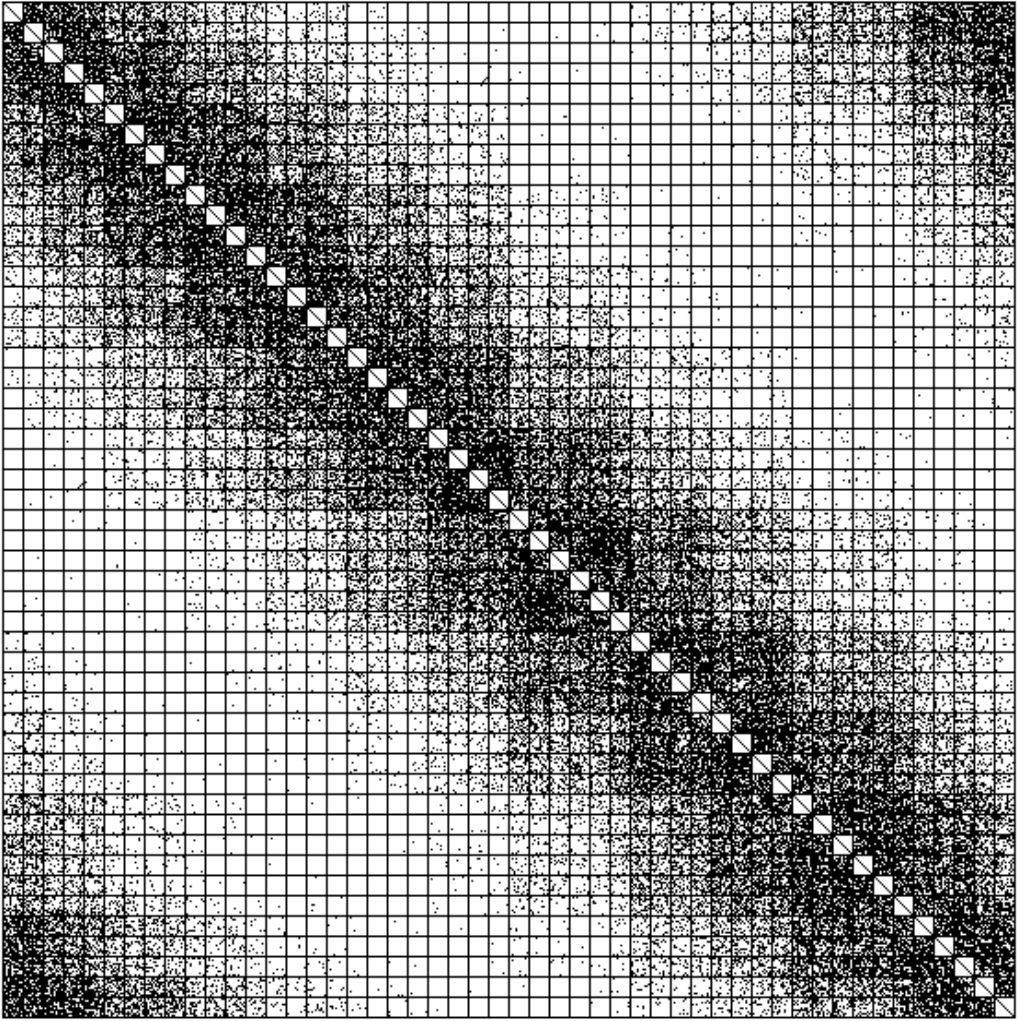
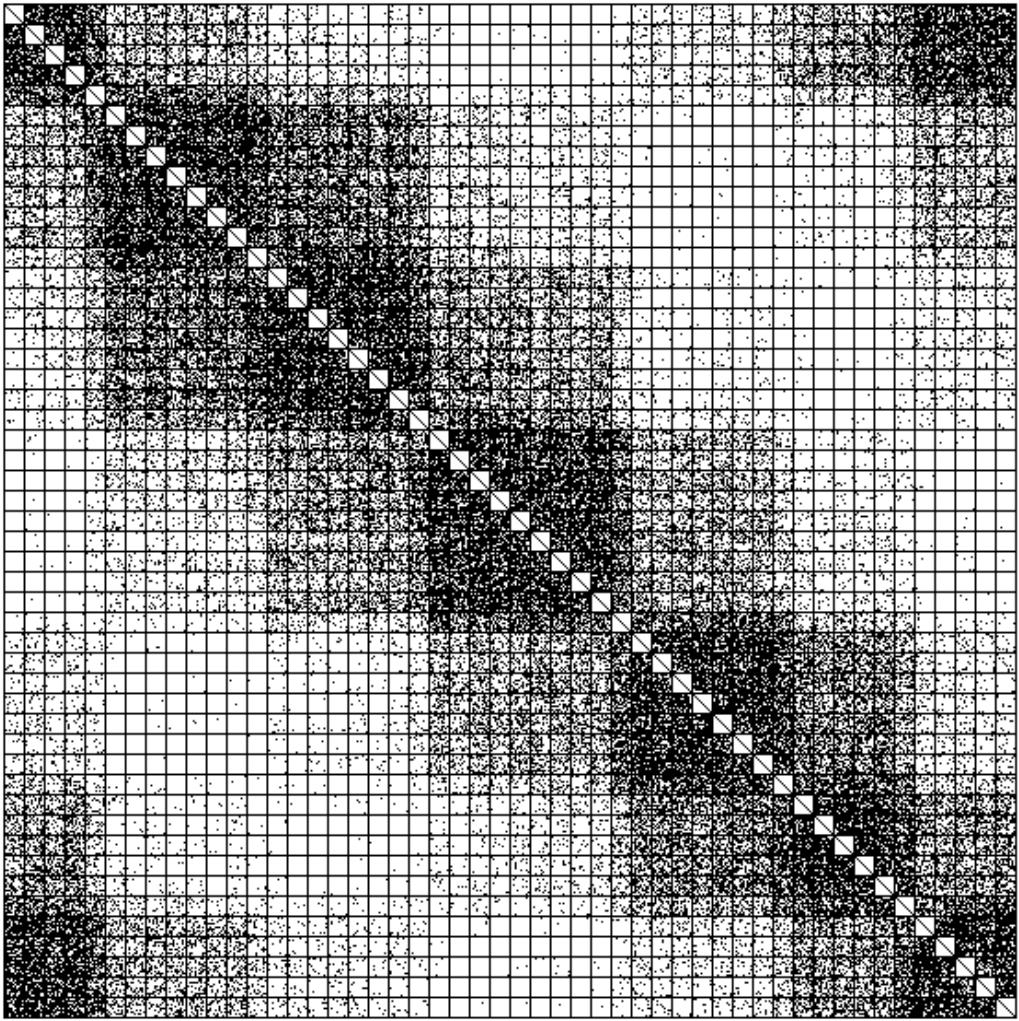
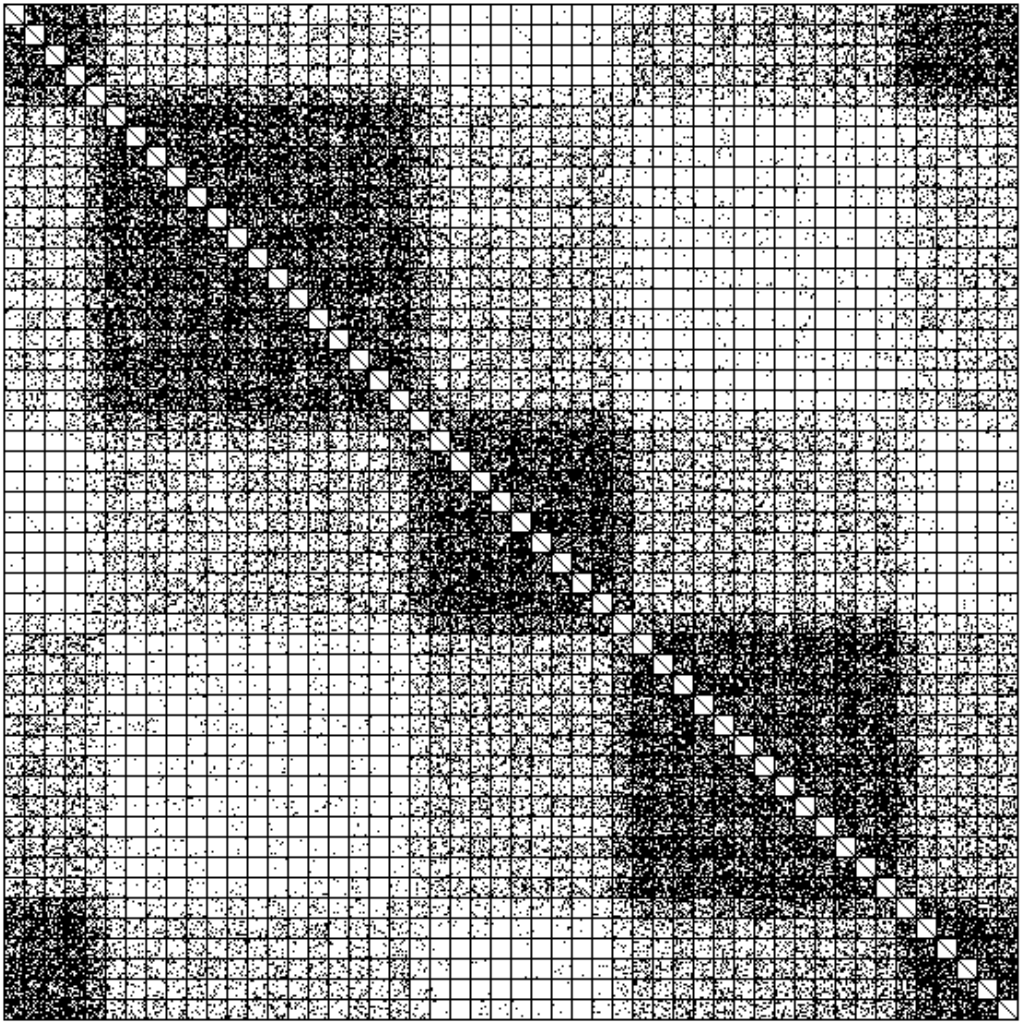
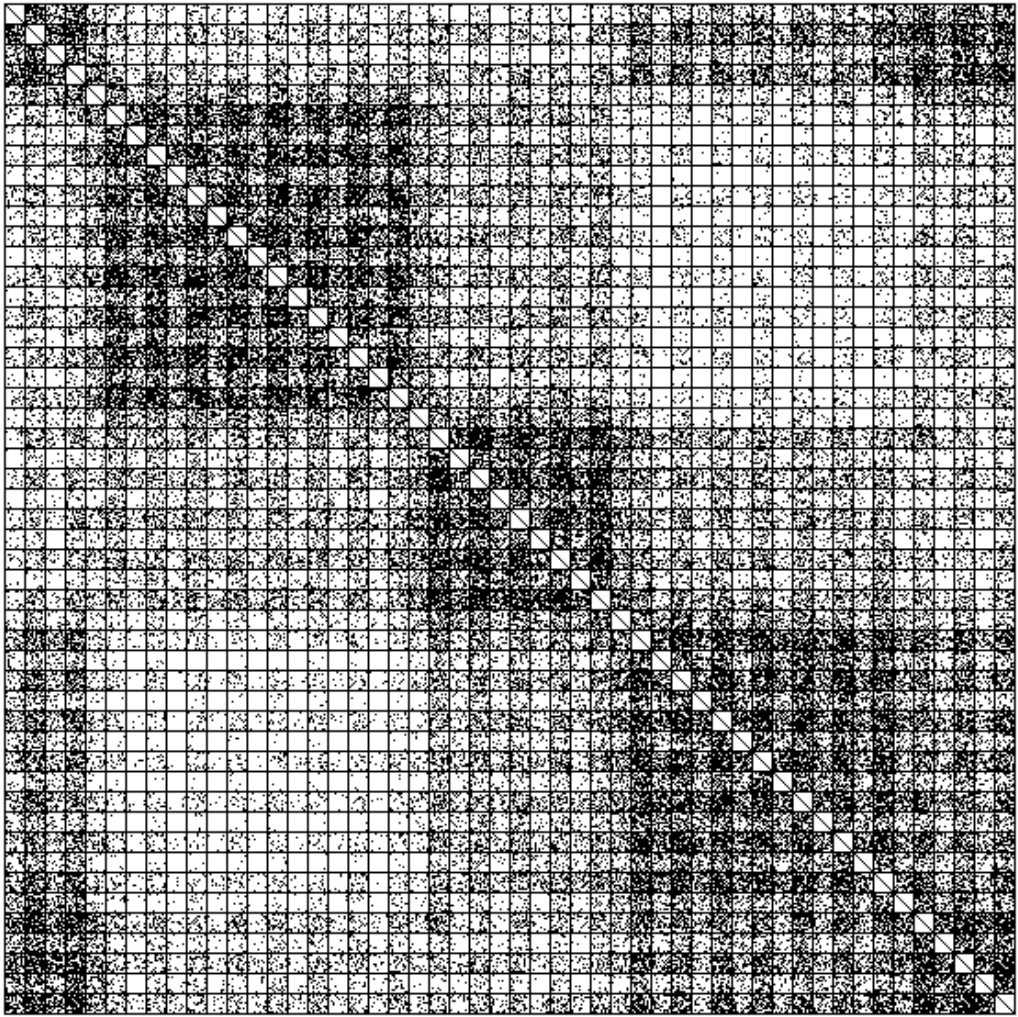
\end{document}